\documentclass[twocolumn,aps,prb,superscriptaddress]{revtex4-2}
\usepackage{color}
\usepackage{mathrsfs}
\usepackage{bm}
\usepackage{amsmath}
\usepackage{amssymb}
\usepackage{graphicx}
\usepackage{hyperref}
\usepackage{color}
\usepackage{siunitx}
\allowdisplaybreaks[3]

\makeatletter
\hypersetup{hypertex=true, colorlinks=true, linkcolor=blue, anchorcolor=blue,citecolor=blue}

\usepackage{dcolumn}
\usepackage{bm}
\usepackage{mathrsfs}
\usepackage{amsfonts}
\usepackage{xcolor}
\usepackage{epstopdf}

\begin{document}
\title{Superconductivity in Two-Dimensional Systems with Unconventional Rashba
Bands }
\author{Ran Wang}
\affiliation{Anhui Province Key Laboratory of Low-Energy Quantum Materials and
Devices, High Magnetic Field Laboratory, HFIPS, Chinese Academy
of Sciences, Hefei, Anhui 230031, China}
\affiliation{Science Island Branch of Graduate School, University of Science and
Technology of China, Hefei, Anhui 230026, China}
\author{Jiayang Li}
\affiliation{Anhui Province Key Laboratory of Low-Energy Quantum Materials and
Devices, High Magnetic Field Laboratory, HFIPS, Chinese Academy
of Sciences, Hefei, Anhui 230031, China}
\affiliation{Science Island Branch of Graduate School, University of Science and
Technology of China, Hefei, Anhui 230026, China}
\author{Xinliang Huang}
\affiliation{Anhui Province Key Laboratory of Low-Energy Quantum Materials and
Devices, High Magnetic Field Laboratory, HFIPS, Chinese Academy
of Sciences, Hefei, Anhui 230031, China}
\affiliation{Science Island Branch of Graduate School, University of Science and
Technology of China, Hefei, Anhui 230026, China}
\author{Lichuan Wang}
\affiliation{Anhui Province Key Laboratory of Low-Energy Quantum Materials and
Devices, High Magnetic Field Laboratory, HFIPS, Chinese Academy
of Sciences, Hefei, Anhui 230031, China}
\affiliation{Science Island Branch of Graduate School, University of Science and
Technology of China, Hefei, Anhui 230026, China}
\author{Rui Song}
\affiliation{Science and Technology on Surface Physics and Chemistry Laboratory,
Mianyang, Sichuan 621908, China}\affiliation{Anhui Province Key Laboratory of Low-Energy Quantum Materials and
Devices, High Magnetic Field Laboratory, HFIPS, Chinese Academy
of Sciences, Hefei, Anhui 230031, China}
\author{Ning Hao}
\email{haon@hmfl.ac.cn}
\affiliation{Anhui Province Key Laboratory of Low-Energy Quantum Materials and
Devices, High Magnetic Field Laboratory, HFIPS, Chinese Academy
of Sciences, Hefei, Anhui 230031, China}

\begin{abstract}
In two-dimensional systems with Rashba spin-orbit coupling, it is well-known that superconductivity can exhibit mixed spin-singlet and spin-triplet parity, with the $\boldsymbol{d}$-vector of spin-triplet pairing aligning parallel to the $\boldsymbol{g}$-vector of Rashba spin-orbit coupling. In this work, we propose a model to describe a two-dimensional system with unconventional Rashba bands and investigate its superconducting properties. We demonstrate that the $\boldsymbol{d}$-vector of spin-triplet pairing can be either parallel or perpendicular to the $\boldsymbol{g}$-vector of Rashba spin-orbit coupling, depending on the nature of the pairing interaction. We also propose a junction setup to identify the predominant pairing in such two similar pairing channels. Additionally, our model predicts the emergence of a subleading spin-singlet chiral $p$-wave topological superconducting state. Significantly, we find that such unconventional Rashba bands and superconducting pairings can be realized on the surface of certain superconducting topological materials, such as trigonal layered PtBi$_{2}$ 
\end{abstract}
\maketitle

\section{Introduction}
In systems characterized by two-dimensional (2D) geometry and Rashba spin-orbit coupling (SOC), the breaking of inversion symmetry leads to the lifting of spin degeneracy, resulting in superconductivity featuring a mixture of singlet and triplet pairings\cite{PhysRevLett.87.037004}. For the superconducting transition temperature $T_{c}$ to be enhanced, the $\boldsymbol{d}$-vector for spin-triplet pairing prefers to align parallel to the $\boldsymbol{g}$-vector associated with Rashba SOC\cite{PhysRevLett.92.097001}. These established principles guide the exploration of unconventional superconductivity not only in complex noncentrosymmetric systems\cite{PhysRevB.107.094507,Xi2016}, such as half-Heusler compands\cite{PhysRevB.93.205138,PhysRevX.8.011029,PhysRevResearch.4.L012017} and transition metal dichalcogenides (TMDs)\cite{PhysRevLett.113.097001,PhysRevB.93.180501,Saito2016}, but also in systems lacking local inversion centers\cite{PhysRevB.84.184533,PhysRevB.86.100507,doi:10.1146/annurev-conmatphys-040521-042511}, such as layered TMDs\cite{PhysRevLett.118.087001,PhysRevB.105.L100506,PhysRevResearch.5.043122}.

\begin{figure}[htbp]
\begin{center}
\includegraphics[width=1.0\columnwidth]{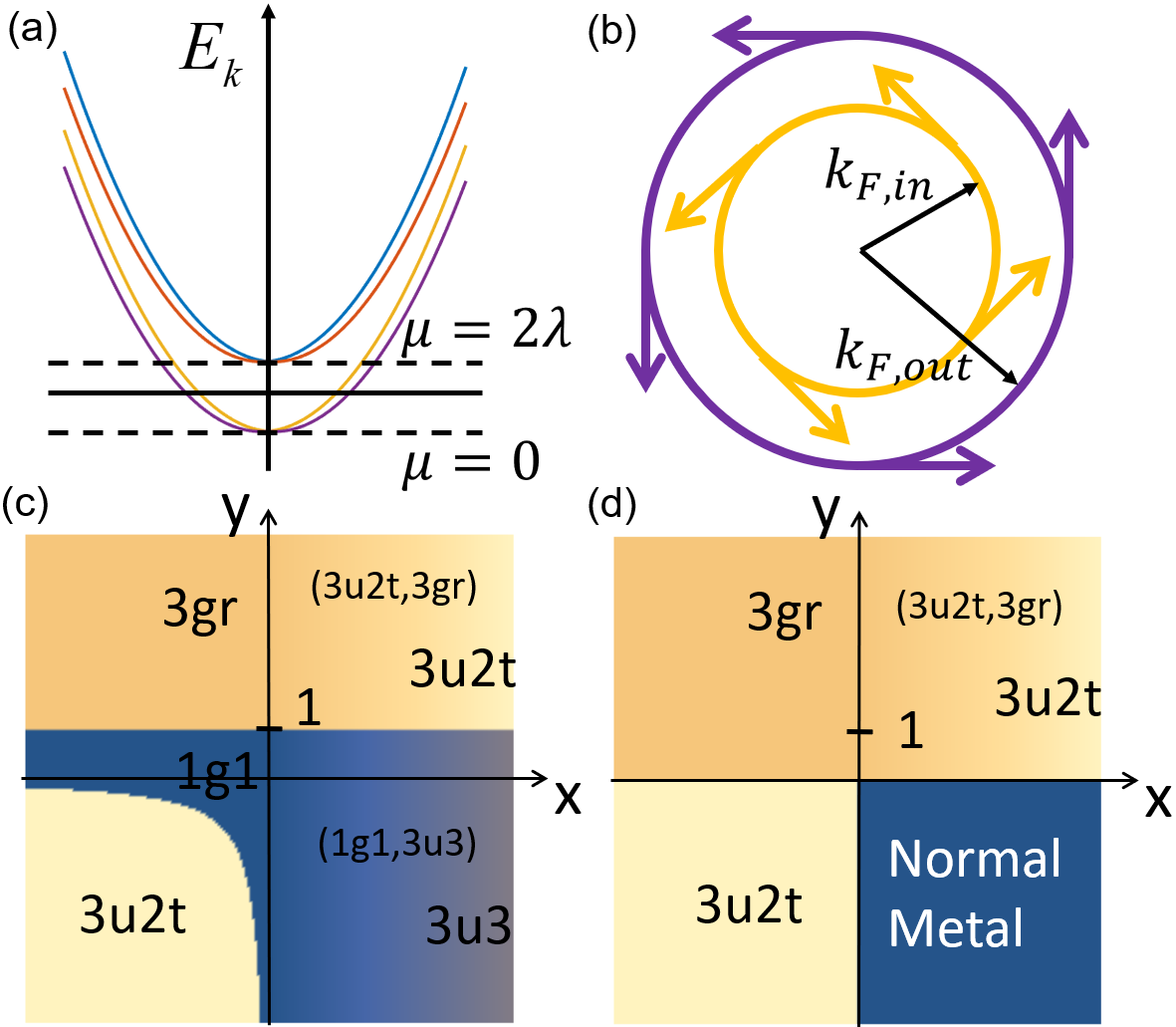}
\end{center}  
\caption{
(a) Energy dispersion of the unconventional Rashba model. The revised chemical potential satisfies $0<\mu<2\lambda$. The splitting between the two lower bands is $\varepsilon\lambda_{R}k$. (b) Spin texture of the two Fermi circles with $\mu=\lambda$ as shown in (a). The Fermi momenta are $k_{F,in(out)}\approx(\sqrt{\varepsilon^2\lambda_{R}^2+4t'\mu}-(+)\varepsilon\lambda_{R})/2t'$. Here, $t'=t-\lambda_{R}^2/2\lambda$. (c) Phase diagram about possible pairings according to their $T_c$ without an external magnetic field. (d) Revised phase diagram under a $z$-direction external magnetic field. $x,y$ are defined by $x=u_{0}/v_{0}=u/v$ and $y=v/v_{0}=u/u_{0}$. Here, $v_{0}$ $(u_{0})$ denotes intra-orbital (inter-orbital) pairing interaction in the zeroth order of $k$. $v$ $(u)$ denotes intra-orbital (inter-orbital) pairing interaction in the first order of $k$. Notably, although we do not constrain $x$ and $y$ in the figures, it is hardly possible to reach the limits of $x(y)\to0(\pm\infty)$ in any real material platform. }
\label{fig1}
\end{figure}

In the two-band Rashba SOC model, Rashba SOC significantly impacts two key aspects. Firstly, it enables the two Fermi circles originating from distinct bands to possess opposing chiral in-plane spin textures. Secondly, it mandates the alignment of the $\boldsymbol{d}$-vector for spin-triplet pairing parallel to the $\boldsymbol{g}$-vector. These aspects are mutually consistent, leading to the conclusion that the $\boldsymbol{g}$-vector of Rashba SOC, the spin texture of the Fermi surface, and the $\boldsymbol{d}$-vector for spin-triplet pairing are all chiral. However, some 2D systems exhibit unconventional Rashba bands\cite{PhysRevB.79.245428,PhysRevB.79.241408,PhysRevB.94.041302,PhysRevB.100.115432}, argued to originate from interband SOC\cite{PhysRevLett.108.196801}, where the two Fermi circles from distinct bands possess identical chiral in-plane spin textures, as depicted in Fig. \ref{fig1}(b). The immediate question that arises is the nature of the superconducting state in such 2D systems with unconventional Rashba bands.

In this work, we introduce a generic model describing a 2D system with unconventional Rashba bands, which can be realized in various material systems. The superconducting pairings are categorized based on the $C_{\infty v}$ group symmetry. By solving the linear gap equation\cite{PhysRevLett.92.097001,PhysRevLett.118.087001}, we identify both mixed singlet and triplet pairings. Specifically, we find two competing spin-triplet pairings: one with its vector parallel and the other perpendicular to the Rashba SOC vector. The superconducting ground state is determined by the dominant pairing interaction. To differentiate between these similar spin-triplet pairings, we propose a metal-superconductor junction. The tunneling currents may exhibit difference due to the distinct quasi-particles of the states. Furthermore, we examine the pairing breaking induced by the Zeeman effect and suggest that our model and findings could be realized on the surface of the trigonal-layered material PtBi$_{2}$, recently reported to exhibit exotic surface superconductivity\cite{Kuibarov2024}. 

This paper is organized as follows. In Sec.\ref{two}, we introduce the unconventional Rashba model and investigate its superconducting properties. In Sec.\ref{three}, we propose a metal-superconductor junction to identify the dominant pairing channel, and further discuss the potential for realizing the unconventional Rashba system in materials. Finally, in Sec.\ref{conclusion}, we give discussions and summarize our results.

\section{Model and Superconductivity of Unconventional Rashba Bands}{\label{two}}
\subsection{Model}
We start from a 4$\times$4 Hamiltonian
$H_{0}$ that yields unconventional Rashba bands\cite{huang2024generic}. In the basis $\hat{\psi}_{\bm{k}}=(c_{\bm{k},1\uparrow},c_{\bm{k},1\downarrow},c_{\bm{k},2\uparrow},c_{\bm{k},2\downarrow})^{T},$
\begin{equation}
	H_{0}(\boldsymbol{k})=\xi_{\boldsymbol{k}}\sigma^{0}\tau^{0}-\lambda_{R}(k_{y}\sigma^{1}-k_{x}\sigma^{2})(\tau^{0}+\varepsilon\tau^{1})+\lambda\sigma^{3}\tau^{2}.\label{H0}
\end{equation}
Here, $i=$1,2 represents effective orbitals or sublattices depending on
the specific material, while $\sigma=\uparrow,\downarrow$ denotes the spin
in the annihilation operator $c_{\bm{k},i\sigma}$. The Pauli matrices
$(\sigma^{0},\bm{\sigma})$ and $(\tau^{0},\bm{\tau})$ span the spin
and orbital spaces respectively. The term $\xi_{\boldsymbol{k}}=tk^2-\mu_0$. $t$ represents the band curvature
nearing Fermi surface. $\mu_0$ is the chemical potential. For convenience, we define a revised chemical potential $\mu=\mu_0+\lambda$, leading to a lower-two-band crossing Fermi surface at $\mu=0$. The term $\lambda_{R}$ describes the intensity of Rashba SOC, while $\varepsilon$ is the ratio between inter- and intra-orbital Rashba SOC. The parameter $\lambda$ stands for on-site SOC.
Explicity, the vector of Rashba SOC is expressed as ${\normalcolor \boldsymbol{g}\mathrm{=(-\mathit{\hat{k}_{y}},\mathit{\hat{k}_{x}},\mathit{\mathrm{0}})}}.$ 
Considering that the two orbitals are related by a spatial inversion operation, specifically, the inversion operation on $H_0=h_{\sigma}(\boldsymbol{k})h_{\tau}$ transforms as $h_{\sigma}(\boldsymbol{k})h_{\tau}\to h_{\sigma}(-\boldsymbol{k})\tau^1h_{\tau}\tau^1$. As a result, both the second and third terms in Hamiltonian (\ref{H0}) break spatial inversion symmetry. $H_0$ exhibits full symmetry of the $C_{\infty v}$ group, which includes continuous rotation about the $z$ axis and infinite mirrors parallel to the $z$ axis. The energy dispersion of $H_0$ is straightforwardly
obtained, 
\begin{equation}
	E_{N,\alpha\beta}=\xi_{\boldsymbol{k}}+\alpha\sqrt{\lambda^{2}+\lambda_{R}^{2}k^{2}}+\beta\varepsilon\lambda_{R}k \label{En}
\end{equation}
Here, $\alpha,\beta\in\{+,-\}$ label the band indices. The spin texture
for band $\alpha$ is $\bm{S}_{\alpha\beta}=\alpha(-k_{y}\bm{e}_{x}+k_{x}\bm{e}_{y})\lambda_{R}/\sqrt{\lambda^{2}+\lambda_{R}^{2}k^{2}}$,
which is along the tangential direction of the Fermi circles for all
four bands. The chirality of the spin textures is the same for the upper
two bands and lower two bands independently while being opposite between
the upper and lower bands, as $\bm{S}_{\alpha\beta}$ is independent
of $\beta$. We obtain two Fermi circles with the same spin chirality if the chemical potential is within the band gap. This corresponds to
unconventional Rashba bands in comparison with conventional Rashba
bands. 

\begin{table}[ptb]
\caption{$\phi_{g/u}$ and $\boldsymbol{d_{g/u}}$ are scalar
and vector functions of $k$, respectively, where g and u labels even
and odd under transformation $\bm{k}\to-\bm{k}$. The second column
lists $\phi_{g/u}$ and $\boldsymbol{d_{g/u}}$ for the simplest pairings to the lowest order of $k$, normalized by $\hat{k}_{x}^{2}+\hat{k}_{y}^{2}=1$
is normalized. The corresponding IRs for $C_{\infty v}$ are listed in
the third column. Pairings denoted in the same font format can mix. In the last column, each pairing is labeled for
convenience: the first number denotes singlet or triplet in spin space,
and the letter g (u) denotes even (odd) behavior under inversion
transformation. The last letters indicate radial states (r), helical
states (t), out-of-plane states (p), and in-plane two-dimensional states (e). }
\label{pair}
\begin{ruledtabular}
\begin{tabular}{cccc}
 Pairing form  & $\phi_{g/u}$ / $\boldsymbol{d_{g/u}}$   & IRs for $C_{\infty v}$  & Label\tabularnewline
\hline 
 $i\phi_{g}\sigma^{2}\tau^{0}$  & $1$  & $A_1$  & 1g1 \tabularnewline
 $i\phi_{g}\sigma^{2}\tau^{1}$  & $1$  & $\mathbf{A_1}$  & 1g2\tabularnewline
 $i\phi_{g}\sigma^{2}\tau^{3}$  & $1$  & $A_2$  & 1u \tabularnewline
 $i\phi_{u}\sigma^{2}\tau^{2}$  & $\hat{k}_x,\hat{k}_y$  & $E_1$  & 1g3 \tabularnewline
 $i(\boldsymbol{d_{u}}\cdot\boldsymbol{\sigma})\sigma^{2}\tau^{0}$  & $(-\hat{k}_{y},\hat{k}_{x},0)$  & $A_1$  & 3u1t \tabularnewline
 & $(0,0,\hat{k}_{x}),(0,0,\hat{k}_{y})$  & $E_1$  & 3u1p \tabularnewline
 & $(\hat{k}_x,-\hat{k}_y,0),(\hat{k}_y,\hat{k}_x,0)$ & $E_2$ &3u1e \tabularnewline
 $i(\boldsymbol{d_{u}}\cdot\boldsymbol{\sigma})\sigma^{2}\tau^{1}$  & $(\hat{k}_{x},\hat{k}_{y},0)$  & $A_2$  & 3u2r \tabularnewline
 & $(-\hat{k}_{y},\hat{k}_{x},0)$  & $\mathbf{A_1}$  & 3u2t \tabularnewline
 & $(\hat{k}_x,-\hat{k}_y,0),(\hat{k}_y,\hat{k}_x,0)$ & $\mathbf{E_2}$ &3u2e \tabularnewline
 $i(\boldsymbol{d_{u}}\cdot\boldsymbol{\sigma})\sigma^{2}\tau^{3}$  & $(\hat{k}_{x},\hat{k}_{y},0)$  & $\mathbf{A_1}$  & 3gr \tabularnewline
 & $(-\hat{k}_{y},\hat{k}_{x},0)$  & $A_2$  & 3gt \tabularnewline
 & $(\hat{k}_y,\hat{k}_x,0),(\hat{k}_x,-\hat{k}_y,0)$ & $\mathbf{E_2}$ &3ge \tabularnewline
 $i(\boldsymbol{d_{g}}\cdot\boldsymbol{\sigma})\sigma^{2}\tau^{2}$  & $(0,0,1)$  & $A_1$  & 3u3 \tabularnewline
 & $(0,1,0),(1,0,0)$ & $E_1$ &3u3e \tabularnewline
\end{tabular}
\end{ruledtabular}
\end{table}

\subsection{Pairing classification}
Before solving the superconductivity
problem specifically, we classify the superconducting pairings
according to the symmetry of Hamiltonian (\ref{H0}). We focus on the case where the splitting of the lower two bands is sufficiently small, such that finite momentum pairings and the difference in the density of states between the two
bands can be neglected. Superconducting pairs are expressed in the basis of creation and annihilation operators $(\hat{\psi}_{\bm{k}},\hat{\psi}_{\bm{k}}^{\dagger})$
as follows: 
\begin{align}
\hat{B}_{\bm{k}} & =(\hat{\psi}_{-\bm{k}})^{T}\mathscr{T}_{\Gamma}^{\dagger}(\bm{k})\hat{\psi}_{\bm{k}},\nonumber \\
\hat{B}_{\bm{k}}^{\dagger} & =\hat{\psi}_{\bm{k}}^{\dagger}\mathscr{T}_{\Gamma}(\bm{k})(\hat{\psi}_{-\bm{k}}^{\dagger})^{T}.
\end{align}
This form is useful in the phenomenological theory of unconventional superconductivity\cite{RevModPhys.63.239,PhysRevB.40.4329,PhysRevB.31.7114}. It is convenient to divide the pairing matrix into two parts, $\mathscr{T}_{\Gamma}(\bm{k})=\gamma(\bm{k})\mathcal{T}$,
where $\mathcal{T}$ is constructed using $\tau^{\mu}$ in orbital space, while $\gamma(\bm{k})$ is
in spin space. 
The momentum dependence is incorporated into $\gamma(\bm{k})$ by constructing
it with power functions of $k$ (basis functions of irreducible representations
(IRs) for group $SO(3)$) and $\boldsymbol{\sigma}$ matrices. Note that the conventional Rashba Hamiltonian also has $C_{\infty v}$ symmetry. To provide an explicit comparison, we classify the pairings of our model according to the $C_{\infty v}$ group. The irreducible pairings for the $C_{\infty v}$ group are listed in Table~\ref{pair}. We only consider the pairings in the zeroth and first order of $k$, i.e., one-dimensional IRs $A_1$, $A_2$, and two-dimensional IRs $E_1$, $E_2$ for $C_{\infty v}$ group. The pairings within the same IR may mix, whereas those in different IRs will not mix. The explicit mixings are determined in the next subsection. Furthermore, the reduction of symmetry can induce mixing and combine several distinct IRs into a single new IR. When a magnetic field is applied in the $z$-direction, the symmetry is reduced from $C_{\infty v}$ to $SO(2)$, with new classifications shown in Appendix \ref{app_field}. For specific materials with an $n$-fold rotational axis, the $C_{\infty v}$ reduces to $C_{nv}$, making the relevant pairing classification straightforward.

\subsection{Solutions of linear gap equation}
According to the phenomenological theory of superconductivity, the pairing interaction $H_{int}$ can be expressed in terms of IRs of the system's symmetry group as follows: 
\begin{equation}
H_{int}=-\frac{1}{2}\sum_{\Gamma,j,\bm{k,k'}}V_{\Gamma,j}\hat{\psi}_{\bm{k}}^{\dagger}\mathscr{T}_{\Gamma,j}(\bm{k})(\hat{\psi}_{-\bm{k}}^{\dagger})^{T}(\hat{\psi}_{-\bm{k'}})^{T}\mathscr{T}_{\Gamma,j}^{\dagger}(\bm{k'})\hat{\psi}_{\bm{k'}}
\end{equation}
Here, $j$ labels the superconducting pairings of IR $\Gamma$, and $V_{\Gamma,j}$ is the parameter of $j$th channel in IR $\Gamma$. This interaction leads to a linear gap equation: 
\begin{equation}
\Delta_{\Gamma,j}=V_{\Gamma,j}\sum_{l}\chi_{\Gamma,jl}\Delta_{\Gamma,l}.\label{gap eq}
\end{equation}
$\Delta_{\Gamma,j}=-V_{\Gamma,j}\sum_{\bm{k}\mu\nu}\langle c_{-\bm{k'}\mu}(\mathscr{T}_{\Gamma,j}^{\dagger})^{\mu\nu}c_{\bm{k'}\nu}\rangle$
labels the superconducting order parameter. $\mu,\nu$ denote both orbital and spin degrees
of freedom. $\chi_{\Gamma,jl}$ is defined
as superconductivity susceptibility between $j$ and $l$ pairings
of IR $\Gamma$:
\begin{equation}
\chi_{\Gamma,jl}=\frac{1}{\beta}\sum_{n,\bm{k}}Tr[\mathscr{T}_{\Gamma,j}^{\dagger}(\bm{k})G^{0}(\bm{k},i\omega_{n})\mathscr{T}_{\Gamma,l}(\bm{k})(G^{0}(-\bm{k},-i\omega_{n}))^{T}]. \label{sus}
\end{equation}
Here, $G^{0}(\bm{k},i\omega_{n})$ is the Matsubara Green function
for the normal state(See Appendix \ref{app_G0} for details), and $\beta=1/k_{B}T$. One can examine the mixing
of states from the linear gap equation(\ref{gap eq}) and obtain critical
temperatures. We consider the case of unconventional Rashba bands
with the chemical potential lying in the gap between the upper two and lower
two bands, as shown in Fig. \ref{fig1}(a). The parameters $V_{\Gamma,j}$ can
differ between s-wave and p-wave pairings as well as between intra- and
inter-orbital pairings. Thus, we introduce four interaction parameters
$V_{\Gamma,j}$: $v_{0}$ for 1g1 and 1u (s-wave, intra-orbital); $u_{0}$ for 1g2 and 3u3 (s-wave, inter-orbital);$v$
for 3u1 and 3g (p-wave, intra-orbital); $u$ for 1g3 and 3u2 (p-wave, inter-orbital), keeping in mind that a nagetive
$V_{\Gamma,j}$ would result in $\Delta_{\Gamma,j}=0$. Interestingly, such kinds of attractive interactions can be induced by spin and charge fluctuation in weak-coupling limit.

From the solutions of Eq.(\ref{gap eq}), there are five types of
separately mixed states: a$\in$\{1g1, 3u1t, 3u3\}; b$\in$\{1g2,
3u2t, 3gr\}; c$\in$\{1u, 3u2r, 3gt\}; d$\in$\{1g3, 3u1p, 3u3e\}; e$\in$\{3u2e, 3ge\}, each
of these states is associated with a critical temperature determined by the
linear gap equation Eq.(\ref{gap eq})(See Appendix \ref{app_phase} for explicit relations). We obtain the phase diagram at critical temperature in Fig.\ref{fig1}(c) by comparing these $T_c$s, which is correct except in the limit $\varepsilon\to0$. Channels 1g2, 1u, 3u1t, 3u1e and 3u3e are excluded under the limit $\lambda_Rk_F\ll\lambda$. 

\begin{figure}[htbp]
	\begin{center}
		\includegraphics[width=1.0\columnwidth]{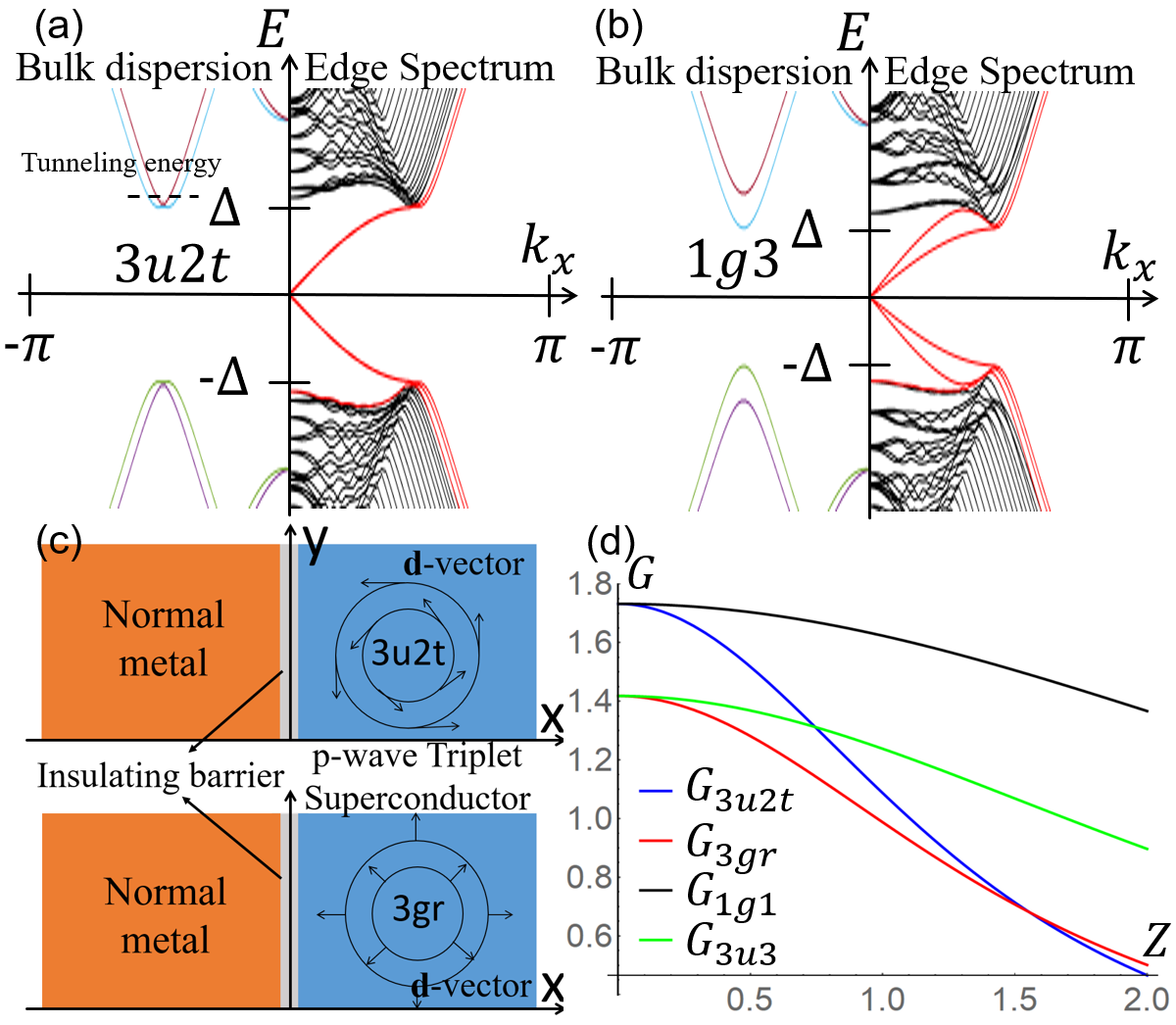}
	\end{center}
	\caption{(a) Bulk (edge) quasiparticle dispersion (spectrum) for the superconducting states with 3u2t pairing. The plotting is similar for 3gr pairing. (b) Bulk (edge) quasiparticle dispersion (spectrum) for superconducting states with 1g3 pairing. The minimized bulk energy in (a) and (b) is around $\Delta$. (c) Schematic diagram of a junction including a normal metal and a unconventional Rashba superconductor with schematic 3u2t pairing or 3gr pairing. (d) Plot of $G$ as a function of $Z$ for junctions with four possible pairings: s-wave \{1g1, 3u3\} and p-wave \{3u2t, 3gr\}. Here, $Z=2mU_0/\hbar^2k_F$. Parameters: $\lambda_Rk_F/2\lambda=0.2$, $\Delta/E=0.95$. } \label{fig-property}
\end{figure}

\subsection{Properties of possible pairings}
The four
possible predominant pairings exhibit different properties. The 1g1 pairing is a
standard spin-singlet pairing, while the 3u3 pairing is a k-independent spin-triplet
pairing. The 3u2t and 3gr pairings
have more intriguing properties, as both yield topological superconducting states characterized by a $Z_{2}=1$ number. Their bulk dispersion and edge spectrum are similar, as illustrated ub Fig.\ref{fig-property} (a) for the 3u2t case. 
Despite these similarities and their ability to mix, their $\boldsymbol{d}$-vectors
are distinctly different: $\mathbf{\mathit{\boldsymbol{d_{3u2t}}}}\parallel\mathbf{\mathit{\boldsymbol{g}}}$
and $\mathbf{\mathit{\boldsymbol{d_{3gr}}}}\perp\mathbf{\mathit{\boldsymbol{g}}}$.
Further, although the 1g3 pairing is subleading and does not appear in the phase
diagram, its $T_{c}$ is comparable to the leading pairing from the
linear gap equation calculation, making it a viable possibility in real materials with complex parameters. When an out-of-plane magnetic field is applied or through spontaneous time-reversal-symmetry-breaking effects, the 1g3 pairing results in spin-singlet chiral $p$-wave
topological superconducting states, characterized by Chern number of $C=\pm2$. The bulk dispersion and edge spectrum for 1g3 are shown in Fig. \ref{fig-property} (b). 

Applying an external magnetic field provides a valuable means to distinguish between pairings in different states \cite{PhysRevLett.92.097001,Saito2016,Xi2016}. These pairs exhibit greater sensitivity to out-of-plane fields due to the higher-order pair-breaking terms introduced by in-plane fields, consistent with Ref.\cite{doi:10.1021/acs.nanolett.2c04297}. Under an out-of-plane magnetic field, the symmetry is reduced from $C_{\infty v}$ to $SO(2)$, which combines states a and c from previous section. The revised linear gap equations are solved(refer to Appendix \ref{app_field} for details), revealing that both the 3u2t and 3gr pairings remain unaffected, while the 1g1 and 3u3 pairings can be suppressed by the field. This suppression can be significant even with a weak field due to both field-induced and symmetry-breaking-induced pair breaking. Other subleading states not in the phase diagram Fig.\ref{fig1}(c) are also suppressed by the field, providing a method to tune different pairings. The revised phase diagram under magnetic field conditions is illustrated in Fig. \ref{fig1}(d). 

\section{Experimental conception and Material Platform}{\label{three}}
\subsection{Identification of predominant pairing with a junction}
We consider all pairing interactions arise from the same mechanism, with both $u/v$ and $u_0/v_0$ having same tendency to be other larger or smaller than 1. This indicates that the predominant pairing is other inter-orbital or intra-orbital type. Furhermore, the phase diagram in Fig. \ref{fig1} indicates the four state mixture of s-wave pairings \{1g1, 3u3\} and p-wave pairings \{3u2t, 3gr\} below a critical temperature due to their same $A_1$ IR. Then, experimentally identifying the predominant pairing becomes possible due to a slight difference in the quasi-particle energy dispersions of these pairings described by Eq.(\ref{3u2t}),(\ref{3gr}). To address this, we propose a normal metal-superconductor junction\cite{PhysRevLett.74.3451,SatoshiKashiwaya2000,CHENG201552,PhysRevLett.108.077002,PhysRevB.93.195404} to measure the tunneling conductance. This approach can amplify the inherent differences in quasi-particle energy dispersions, which are given by:
\begin{align}
	E_{S,\alpha\beta}^{1g1/3u2t}&=\pm\sqrt{\left(\xi_{\boldsymbol{k}}+\alpha\sqrt{\lambda^{2}+\lambda_{R}^{2}k^{2}}+\beta\varepsilon\lambda_{R}k\right)^2+\Delta^2} \label{3u2t}\\
	E_{S,\alpha\beta}^{3u3/3gr}&=\pm\sqrt{\left(\alpha \sqrt{\lambda^2+\lambda_R^2k^2}+f_{\beta}\right)^2+\Delta^2cos^2\phi} \label{3gr}.
\end{align}
Here, $\alpha,\beta$ are as defined in Eq. (\ref{En}). $\phi=arctan(\lambda_{R}k/\lambda)$, $f_{\beta}=\sqrt{(\xi_{\boldsymbol{k}}+\beta\varepsilon\lambda_Rk)^2+\Delta^2sin^2\phi}$. The superconducting order parameter $\Delta$ is an arbitrary real constant. Notably, the ratio $\Delta/E$ (where $E$ is the tunneling energy) rather than magnitude of $\Delta$ plays a crucial role in determining the tunnelling conductance. This feature makes the experiment feasible without requiring identical gap amplidudes when measuring different pairings. 

The junction lies in $x-y$ plane, featuring a barrier at $x=0$ that divides two distinct regions, as schematically depicted in Fig. \ref{fig-property} (c). The normal metal occupies the region $x<0$, while the unconventional Rashba superconductor redises in the region where $x>0$. To demonstrate the conductance differences without complicated calculations, we employ an effective Hamiltonian for the lower two bands(See Appendix \ref{app_tunnel} for details). The Hamiltonian describing the junction is expressed as
\begin{equation}
H_t=(\xi_{\boldsymbol{k}}-\lambda)\sigma^0\eta^3\Theta(-x)+\hat{H}_{BdG}^{eff}\Theta(x)+U_0\sigma^0\eta^3\delta(x). \label{junction}
\end{equation}
Here, the Pauli matrices $(\eta^0,\boldsymbol{\eta})$ span the particle-hole space, and $\xi_{\boldsymbol{k}}$ is identical to that in Eq. (\ref{H0}). $U_0$ denotes barrier potential.
For simplicity, we assume equal effective mass and chemical
potential throughout the junction. $\delta(x)$ represents the delta
function, and $\Theta(x)$ denotes the step function. The Hamiltonian in Eq. (\ref{junction})
can be solved under the continuity condition of the wave function at the
interface (see Appendix \ref{app_tunnel} for details). The conductance as a function of barrier intensity is plotted in Fig. \ref{fig-property}(d). 
In the limit of zero barrier $Z=0$, We find the conductance of the 3u2t pairing equals that of the 1g1 pairing ($G_{1}=G_{1g1}=G_{3u2t}$), while the conductance of the 3gr pairing equals that of the 3u3 pairing ($G_{2}=G_{3u3}=G_{3gr}$). The difference in conductance is given by:
\begin{equation}
	\delta G=G_1-G_2\approx2\left(\frac{\lambda_Rk_F}{2\lambda}\right)^2\frac{(\Delta/E)^2}{\sqrt{1-(\Delta/E)^2}}. \label{G}
\end{equation}
Here, $E=eV_b$ with $V_b$ being experimentally tunable bias voltage. From Eq. (\ref{G}), we find that $\delta G$ is enhanced when $E$ slightly exceeds $\Delta$. Significantly, this enhancement is independent of the relative intensity $\varepsilon$ between inter- and intra-orbital Rashba SOC.

To proceed, prepare a junction with $Z \approx 0$ and start the following detailed experimental measurement at a temperature where s-wave and p-wave pairings are mixed: (i) Use a weak magnetic field to tune the superconducting state into a p-wave pairing state \{3u2t, 3gr\} according to the phase diagram in Fig. \ref{fig1}(d). (ii) Measure the superconducting gap and conduct a tunneling experiment to measure the conductance $G_{p}$ at a fixed ratio $\Delta/E \lesssim 1$. Note that $\Delta$ in Eqs. (\ref{3u2t}, \ref{3gr}) is determined by the minimized energy gap. The factor $\cos^2\phi$ in Eq. (\ref{3gr}) introduces a higher-order term in $(\lambda_R k_F / 2\lambda)$ to Eq. (\ref{G}), which is negligible. (iii) At zero magnetic field, introduce light nonmagnetic dopants to tune the superconducting state into s-wave pairing state \{1g1, 3u3\}. Note that $k$-independent s-wave pairings are robust against nonmagnetic impurities, while $k$-dependent p-wave pairings are sensitive to them \cite{PhysRevLett.75.3938}. (iv) Repeat step (ii) under the conditions of (iii) to obtain a new conductance $G_{s}$.

Using the conductance shown in Fig. \ref{fig-property}(d), deduce a predominant pairing from intra-orbital interaction if $G_{s}>G_{p}$, otherwise deduce predominant one from inter-orbital interaction. Further distinguishing singlet from triplet or s-wave from p-wave pairing is straightforward, such as through upper critical field $H_{c2}$ or topological edge state etc. For predominant pairing lying in p-wave channel, one can identify it as 3gr if $G_{s}>G_{p}$ and as 3u2t if $G_{s}<G_{p}$. Note that the aforementioned method is not suitable under certain limiting conditions, such as $y\gg1$ in Fig. \ref{fig1} (c). In this case, the mixing of s-wave and p-wave pairings cannot be achieved. However, s-wave and p-wave mixed states are common in material platforms lacking inversion symmetry.

\begin{figure}[htbp]
\begin{center}
\includegraphics[width=1.0\columnwidth]{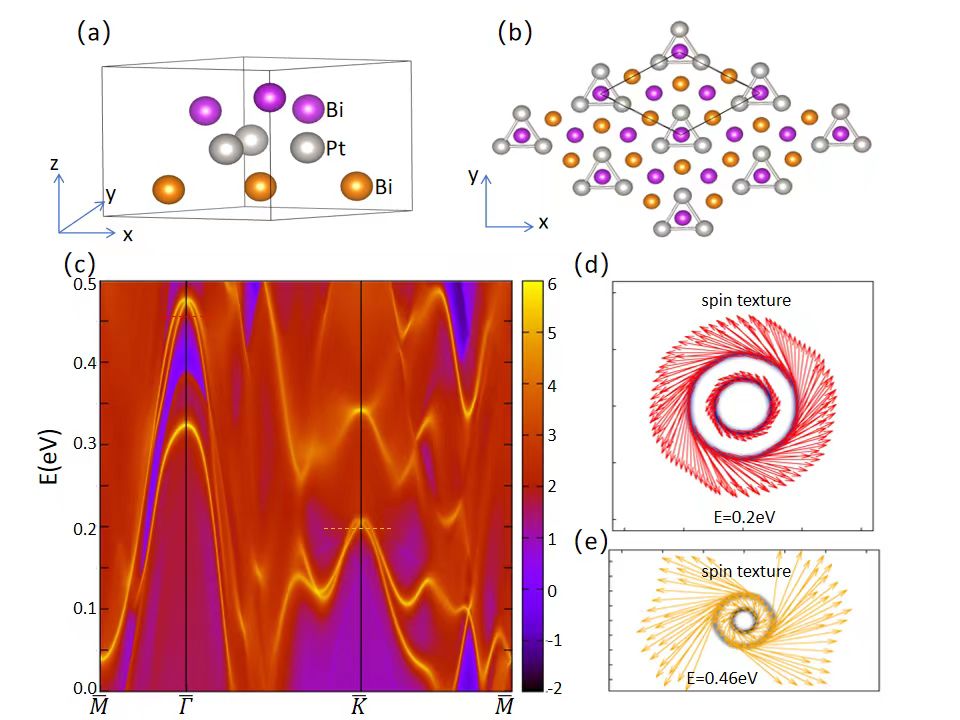}
\end{center}
\caption{(a)-(b) The crystalline structure of trigonal layered PtBi$_{2}$, where the top Bi layer forms a buckled structure breaking inversion symmetry. (c) The Pt cleavage surface spectrum calculated from a slab geometry, plotted along the high-symmetry line in the reduced surface Brillouin zone. (d)-(e) The spin texture for the two Fermi circles at specified filling levels shown in (c).}
\label{fig-PtBi2}
\end{figure}

\subsection{Possible material platform}
We propose that unconventional Rashba bands can be realized on the surface of a trigonal layered PtBi$_{2}$ compound. In our previous work \cite{Gao2018}, we confirmed that trigonal layered PtBi$_{2}$ tends to exhibit a buckled structure for the top Bi layer, as depicted in Fig. \ref{fig-PtBi2} (a)-(b). This structural configuration naturally breaks centrosymmetry and possesses significant Rashba spin-orbit coupling (SOC). Calculations shown in Fig. \ref{fig-PtBi2} (c) demonstrate that the Pt cleavage surface bands exhibit a pronounced spin-splitting feature, and the spin texture under specific doping conditions exhibits the distinct properties of the aforementioned unconventional Rashba bands.
Moreover, this compound has recently been identified as a superconductor below a temperature of several Kelvin \cite{doi:10.1021/acs.nanolett.2c04297, PhysRevMaterials.4.124202}. Consequently, the surface unconventional Rashba bands also exhibit superconducting states, likely through either a bulk-surface proximity effect or effective surface pairing interactions. Recent experiments have observed intriguing phenomena \cite{zabala2023enhanced, schimmel2023hightc, Kuibarov2024} relevant to surface superconductivity. In our scenario, the unconventional Rashba bands can be realized by fine-tuning the doping level through neighboring element replacements. 

\section{Discussion and conclusion}{\label{conclusion}}
In the model Hamiltonian in Eq. (\ref{H0}), the intra-orbital Rashba SOC terms are set to be equal. For the unequal intra-orbital Rashba SOC, an additional $\tau^{3}$ term in Rashba SOC coupling is introduced. Then, the symmetry of $H_{0}$ in Eq. (\ref{H0}) is lowered from $C_{\infty v}$ to $SO(2)$. This reduction of the symmetry would induce further mixed channels similar to the case with out-of-plane magnetic field as we discussed in the paper (See Appendix \ref{app_field}). From the symmetry analysis, we can get the information of which states are mixed, and deduce that the only effect of the $\tau^{3}$ term leads to the possible couplings between \{1g1,3u3\} (the states we are interested in) and \{3u2r,3gt\}. In the limit of this term going to zero, the couplings just mix a tiny component of \{3u2r,3gt\} to \{1g1,3u3\}. It does not change the main picture and present results.

In summary, we introduce an unconventional Rashba band model that significantly departs from the conventional Rashba model. We demonstrate that in our model, the $\boldsymbol{d}$-vector of spin-triplet pairing can align either parallel or perpendicular to the $\boldsymbol{g}$-vector of Rashba spin-orbit coupling, depending on different pairing interactions. This finding is different from the established notion that the $\boldsymbol{d}$-vector must align parallel to the $\boldsymbol{g}$-vector in two-dimensional systems with Rashba spin-orbit coupling.
Furthermore, we propose an experimentally feasible junction to identify and characterize the predominant p-wave pairing state with varying $\boldsymbol{d}$-vectors. Importantly, certain material systems exhibit such unconventional Rashba bands, with the trigonal layered PtBi$_{2}$ showing promising characteristics. Our investigation provides new insights for exploring unconventional superconductivity in non-centrosymmetric materials with unique band structures.

\begin{acknowledgments}
We thank Songbo Zhang for helpful discussions. This work was financially
supported by the National Key R\&D Program of China (Grants No.
2022YFA1403200), National Natural Science Foundation of
China (Grants No. 92265104, No. 12022413, No. 11674331), the Basic Research Program of the Chinese
Academy of Sciences Based on Major Scientific Infrastructures (Grant No. JZHKYPT-2021-08), the CASHIPS Director’s Fund (Grant No. BJPY2023A09), the \textquotedblleft
Strategic Priority Research Program (B)\textquotedblright\ of the Chinese
Academy of Sciences (Grant No. XDB33030100), Anhui Provincial Major S\&T Project(s202305a12020005) and the Major Basic Program of NaturalScience Foundation of Shandong Province (Grant No. ZR2021ZD01), and
the High Magnetic Field Laboratory of Anhui Province under Contract No. AHHM-FX-2020-02.
\end{acknowledgments}

\appendix
\section{Classification of Superconducting Pairs}
We claim in the main text that the pairing matrix is divided into two parts, $\mathscr{T}_{\Gamma}(\bm{k})=\gamma(\bm{k})\mathcal{T}$. Here we are going to elaborate the whole pair parity under inversion based on the character of $\mathcal{T}$ and the symmetry of the system. In our assumption of two obitals being spacial inversion,
one should replace the field $c_{i\sigma}^{\dagger}(\bm{k})$ by $c_{3-i,\sigma}^{\dagger}(-\bm{k})$
as taking inversion operation, which is equivalent to the transformation
on the matrices $\gamma(\bm{k})\mathcal{T}_{ij}\to-\gamma^{T}(\bm{k})(\mathcal{T}^{T})_{3-i,3-j}$
or $\gamma(\bm{k})\mathcal{T}\to-\gamma^{T}(\bm{k})\tau^{1}\mathcal{T}^{*}\tau^{1}$. While the inversion operation on $H_0$ (Eq. (\ref{H0}) in the main text) is $h_\sigma(\boldsymbol{k})h_\tau\to h_\sigma(-\boldsymbol{k})\tau^1h_\tau\tau^1$. In the context of this assumption, the full symmetry of Hamiltonian is group $C_{\infty v}$. Additionally, if assuming the two orbitals tobe different orbitals of atoms, the physical symmetry would be reduced. However, the physical meaning of the basis does not change algebraic structure of $H_0$, thus its implicit symmetry remains an isomorphic group of $C_{\infty v}$, and the results presented in this letter are not affected by the orbital definition. 
\section{\label{app_G0}Matsubara Green function}
The unitary transformation matrix $U$ diagonalizes normal state Hamiltonian $H_0$ (\ref{H0})
\begin{equation}
	U^{-1}H_0U=\begin{pmatrix}
		E_{++}&&&\\
		&E_{+-}&&\\
		&&E_{-+}&\\
		&&&E_{--}
	\end{pmatrix}
\end{equation}
We choose one explicit form
\begin{align}
	&U=\frac{1}{2}[1-isin\theta cos\phi\sigma^1+icos\theta cos\phi\sigma^2+icos\theta\sigma^1\tau^1 \notag \\
	&+isin\theta\sigma^2\tau^1+icos\phi\sigma^3\tau^1-isin\phi\sigma^3\tau^2-sin\theta sin\phi\sigma^1\tau^3 \notag \\
	&+cos\theta sin\phi\sigma^2\tau^3]
\end{align}  
where $\theta=arctan(k_y/k_x),\phi=arctan(\lambda_Rk/\lambda)$. The Matsubara Green function for normal state electrons is
\begin{align}
	G^{0}(\boldsymbol{k},i\omega_n)=U\tilde{G}(\boldsymbol{k},i\omega_n)U^{-1}
\end{align}
here $\tilde{G}$ is Green function for the eigenstates, $\omega_n=(2n+1)\pi/\beta$. 
\begin{equation}
	\tilde{G}(\boldsymbol{k},i\omega_n)=\frac{1}{4}\sum_{\alpha,\beta}\tilde{G}_{\alpha\beta}(\boldsymbol{k},i\omega_n)(\sigma^0+\beta\sigma^3)\otimes(\tau^0+\alpha\tau^3)
\end{equation}
\begin{equation}
	\tilde{G}_{\alpha\beta}(\boldsymbol{k},i\omega_n)=\frac{1}{i\omega_n-E_{\alpha\beta}}
\end{equation}
Then we get Green function for normal state
\begin{align}
	&G^0(\boldsymbol{k},i\omega_n)=G_{++++}\sigma^0\tau^0+G_{+-+-}(\hat{\boldsymbol{k}}\times\boldsymbol{\sigma})_z\tau^1 \notag \\
	&+G_{++--}[\frac{\lambda_Rk}{\sqrt{\lambda^2+\lambda_R^2k^2}}(\hat{\boldsymbol{k}}\times\boldsymbol{\sigma})_z\tau^0+\frac{\lambda}{\sqrt{\lambda^2+\lambda_R^2k^2}}\sigma^3\tau^2] \notag \\
	&+G_{+--+}[\frac{\lambda_Rk}{\sqrt{\lambda^2+\lambda_R^2k^2}}\sigma^0\tau^1-\frac{\lambda}{\sqrt{\lambda^2+\lambda_R^2k^2}}(\hat{\boldsymbol{k}}\cdot\boldsymbol{\sigma})\tau^3] \label{G0}
\end{align}
where $G_{\alpha\beta\gamma\delta}=(\alpha \tilde{G}_{++}+\beta \tilde{G}_{+-}+\gamma \tilde{G}_{-+}+\delta \tilde{G}_{--})/4$.

\section{\label{app_susc}Superconductivity susceptibility}
The calculation of superconductivity susceptibilities(Eq.(\ref{sus}) in the main text) involves Sumation in momentum space, which is handled in the conventional way of changing into integral in energy $\sum_{\boldsymbol{k}}\to \frac{1}{2\pi}N\int d\Omega\int d\epsilon$. $N$ is DOS for each band and $d\Omega$ is the infinitesimal solid angle of 2D. The general results for $\frac{d}{d\Omega}[\frac{1}{\beta}\sum_{n,\boldsymbol{k}}G(\boldsymbol{k},i\omega_n)G'(-\boldsymbol{k},-i\omega_n)]$ are
\begin{align}
	&\frac{d}{d\Omega}\left[-\frac{1}{\beta}\sum_{n,\boldsymbol{k}}\frac{1}{i\omega_n-\epsilon_k}\frac{1}{i\omega_n+\epsilon_k}\right] \notag \\
	&= N\int_{-\omega_D}^{+\omega_D} d\epsilon\frac{\tanh \frac{\beta\epsilon}{2}}{2\epsilon}= N\ln\frac{2e^\gamma\omega_D}{\pi k_BT}\equiv\frac{1}{N_B}\chi_0 \label{chi0} \\
	&\frac{d}{d\Omega}\left[-\frac{1}{\beta}\sum_{n,\boldsymbol{k}}\frac{1}{i\omega_n-\epsilon_k+x}\frac{1}{i\omega_n+\epsilon_k+y}\right] \notag \\
	&= \begin{cases}
		[\frac{1}{N_B}\chi_0+N\mathcal{C}_0(T,\frac{x+y}{2})], (|
		x|,|y|\ll\omega_D) \\
		0, (others)
	\end{cases}
\end{align}
$\epsilon_k$ is energy measured from Fermi surface, $\omega_D$ is Debye frequency.  $\mathcal{C}_0(T,x)=Re[\psi^{(0)}(\frac{1}{2})-\psi^{(0)}(\frac{1}{2}+i\frac{x}{2\pi k_BT})]$, where $\psi^{(0)}(z)$ is the di-gamma funtion. $N_B$ is the number of bands near Fermi surface, only which contribute to superconductivity. $\chi_0$ is the superconductivity susceptibility for the spin singlet state in standard Bardeen-Cooper-Schrieffer
(BCS) theory of 2D. We assume the Rashba SOC is relatively weak campared to the on-site one ($\lambda_Rk_F\ll \lambda$), meanwhile let Fermi surface to stay in the gap of upper and lower bands for the simplicity. Define the notation $g_{\mu\nu\rho\sigma}^{\alpha\beta\gamma\delta}=\frac{d}{d\Omega}[\frac{1}{\beta}\sum_{n,\boldsymbol{k}}G_{\alpha\beta\gamma\delta}(\boldsymbol{k},i\omega_n)G_{\mu\nu\rho\sigma}(-\boldsymbol{k},-i\omega_n)]$, whose  results are
\begin{align}
	g_{++++}^{++++}&=g_{++--}^{++--}=-g_{++--}^{++++}=-g_{++++}^{++--}\notag \\
	&=\frac{1}{8}\chi_0+\frac{1}{8}N\mathcal{C}_0(T,\varepsilon\lambda_Rk_F) \\
	g_{+-+-}^{+-+-}&=g_{+--+}^{+--+}=-g_{+--+}^{+-+-}=-g_{+-+-}^{+--+}\notag \\
	&=-\frac{1}{8}N\mathcal{C}_0(T,\varepsilon\lambda_Rk_F) 
\end{align}
$g_{\mu\nu\rho\sigma}^{\alpha\beta\gamma\delta}$s not listed above are zero. Since we are discussing about pairings to the first order of $k$ and there is dependence of $\hat{k}_i=k_i/|\boldsymbol{k}|$($i=x,y,z$) in Eq.~(\ref{G0}), one could concern about the sumations in form of $\frac{1}{\beta}\sum_{n,\boldsymbol{k}}k_ik_jG(\boldsymbol{k},i\omega_n)G'(-\boldsymbol{k},-i\omega_n)$, etc. However the $\hat{\boldsymbol{k}}$ dependence will just goes into solid angle integral provided  pairings are normalized, thus we will use superconductivity susceptibility in differential sense $\tilde{\chi}_{\Gamma,ij}=d\chi_{\Gamma,ij}/d\Omega$ for convenience. 

\subsection{Singlet states:}
\begin{align}
	\chi_{1g1,1g1}=&\chi_0 \\
	\chi_{1g1,1g2}=&\chi_{1g2,1g1}=0 \\
	\chi_{1g2,1g2}=&\frac{\lambda_R^2k^2}{\lambda^2+\lambda_R^2k^2}\chi_0 \\
	\chi_{1u,1u}=&\frac{\lambda_R^2k^2}{\lambda^2+\lambda_R^2k^2}(\chi_0+2N\mathcal{C}_0(T,\varepsilon\lambda_Rk_F)) \\
	\chi_{1u,1g1}=&\chi_{1g1,1u}=\chi_{1u,1g2}=\chi_{1g2,1u}=0 \\
	\tilde{\chi}_{1g3,1g3}=&|\phi_1|^2[\chi_0+2N\mathcal{C}_0(T,\varepsilon\lambda_Rk_F)]\label{1g3}
\end{align}
1g1 and 1g2 correspond to $A_1$ representation for group $C_{\infty v}$, while 1u corresponds to $A_2$. 1g3 is the first  $\hat{k}_i(i=x,y,z)$ dependent pairing, which corresponds to $E_1$ representation, however the two independent normalized factors $\phi_1=\sqrt{2}\hat{k}_x(\hat{k}_y)$ contributing to a factor $1$ in solid angle integral make $\chi_{1g3,1g3}$ no more than $g_{\mu\nu\rho\sigma}^{\alpha\beta\gamma\delta}$s.
\subsection{Triplet states:}
Calculations all depend on solid angle integral for triplet states. Wirte $\tilde{\chi}$ in form of $\tilde{\chi}_0+\Delta\tilde{\chi}$, $\Delta\tilde{\chi}$ is the pair breaking term which describes the deviation of critical temperature from $T_{c0}$. $\Delta\tilde{\chi}$ is always minimized by radial($\boldsymbol{d}\parallel\hat{\boldsymbol{k}}$), helical($\boldsymbol{d}\perp\hat{\boldsymbol{k}}$), or out of plane($\boldsymbol{d}_{\parallel}=0$) superconducting states. Therefore the states we focus on have maximum superconductivity susceptibilities and are independent of integral in solid angle. The helical state(3u1t) gives out an $A_1$ representation, while the out of plane two-dimensional states(3u1p) are $E_1$ representation, the in plane two-dimensional states(3u1e) are $E_2$ representation. 
\begin{align}
	&\tilde{\chi}_{3u1,3u1}=\frac{\lambda_R^2k_F^2}{\lambda^2+\lambda_R^2k_F^2}(|\boldsymbol{d}_{\parallel}|^2-|\boldsymbol{d}\cdot\hat{\boldsymbol{k}}|^2)\chi_0\notag \\
	&+\frac{\lambda^2}{\lambda^2+\lambda_R^2k_F^2}|\boldsymbol{d}_{\perp}|^2(\chi_0+2N\mathcal{C}_0(T,\varepsilon\lambda_Rk_F)) \notag \\
	\to&\begin{cases}
		&\chi_{3u1t,3u1t}=\frac{\lambda_R^2k_F^2}{\lambda^2+\lambda_R^2k_F^2}\chi_0,\quad\boldsymbol{d}=(-\hat{k}_y,\hat{k}_x,0). \\
		&\chi_{3u1p,3u1p}=\frac{\lambda^2}{\lambda^2+\lambda_R^2k_F^2}(\chi_0+2N\mathcal{C}_0(T,\varepsilon\lambda_Rk_F)), \\
		&\boldsymbol{d}=\sqrt{2}(0,0,\hat{k}_x)/\boldsymbol{d}=\sqrt{2}(0,0,\hat{k}_y). \\
		&\chi_{3u1e,3u1e}=\frac{\lambda_R^2k_F^2}{\lambda^2+\lambda_R^2k_F^2}\chi_0, \\
		&\boldsymbol{d}=\sqrt{2}(\hat{k}_x,-\hat{k}_y,0)/\boldsymbol{d}=\sqrt{2}(\hat{k}_y,\hat{k}_x,0).
	\end{cases} \label{3u1}
\end{align}
3u2 is divided to a radial state(3u2r) and a helical state(3u2t), which corresponds to $A_2$ and $A_1$ representations for group $C_{\infty v}$ respectively, together with a two-dimensional $E_2$ representation(3u2e).
\begin{align}
	&\tilde{\chi}_{3u2,3u2}=\frac{\lambda^2}{\lambda^2+\lambda_R^2k^2}|\boldsymbol{d}\cdot\hat{\boldsymbol{k}}|^22N\mathcal{C}_0(T,\varepsilon\lambda_Rk_F) \notag \\
	&+(\frac{\lambda^2}{\lambda^2+\lambda_R^2k^2}|\boldsymbol{d}_{\parallel}|^2+\frac{\lambda_R^2k^2}{\lambda^2+\lambda_R^2k^2}(|\boldsymbol{d}_{\parallel}|^2-|\boldsymbol{d}\cdot\hat{\boldsymbol{k}}|^2))\chi_0 \notag \\
	&\to\begin{cases}
		&\chi_{3u2r,3u2r}=\frac{\lambda^2}{\lambda^2+\lambda_R^2k_F^2}(\chi_0+2N\mathcal{C}_0(T,\varepsilon\lambda_Rk_F)), \\
		&\boldsymbol{d}=(\hat{k}_x,\hat{k}_y,0). \\
		&\chi_{3u2t,3u2t}=\chi_0,\quad\boldsymbol{d}=(-\hat{k}_y,\hat{k}_x,0). \\
		&\chi_{3u2e,3u2e}=\frac{1}{2}\left(1+\frac{\lambda^2}{\lambda^2+\lambda_R^2k_F^2}\right)\chi_0 \\
		&+\frac{\lambda^2}{\lambda^2+\lambda_R^2k_F^2}N\mathcal{C}_0(T,\varepsilon\lambda_Rk_F), \\
		&\boldsymbol{d}=(\hat{k}_x,-\hat{k}_y,0)/(\hat{k}_y,\hat{k}_x,0).
	\end{cases}
\end{align}
There are also three channels in 3g, radial state(3gr), helical state(3gt) and two-dimensional states(3ge), they correspond to $A_1$, $A_2$ and $E_2$ representations. 
\begin{align}
	&\tilde{\chi}_{3g,3g}=\frac{\lambda^2}{\lambda^2+\lambda_R^2k^2}|\boldsymbol{d}_{\parallel}|^2\chi_0 \notag \\ 
	&+(|\boldsymbol{d}_{\parallel}|^2-|\boldsymbol{d}\cdot\hat{\boldsymbol{k}}|^2)(\frac{\lambda_R^2k^2}{\lambda^2+\lambda_R^2k^2}\chi_0+2N\mathcal{C}_0(T,\varepsilon\lambda_Rk_F))  \notag \\
	&\to\begin{cases}
		&\chi_{3gr,3gr}=\frac{\lambda^2}{\lambda^2+\lambda_R^2k_F^2}\chi_0, \quad\boldsymbol{d}=(\hat{k}_x,\hat{k}_y,0). \\
		&\chi_{3gt,3gt}=\chi_0+2N\mathcal{C}_0(T,\varepsilon\lambda_Rk_F),\quad\boldsymbol{d}=(-\hat{k}_y,\hat{k}_x,0).\\
		&\chi_{3ge,3ge}=\frac{1}{2}\left(1+\frac{\lambda^2}{\lambda^2+\lambda_R^2k_F^2}\right)\chi_0+N\mathcal{C}_0(T,\varepsilon\lambda_Rk_F),\\
		&\boldsymbol{d}=(\hat{k}_y,\hat{k}_x,0)/(\hat{k}_x,-\hat{k}_y,0).
	\end{cases}
\end{align}
For 3u3, there are $A_1$(3u3) and $E_1$(3u3e) states. 
\begin{align}
	&\tilde{\chi}_{3u3,3u3}=\frac{\lambda^2}{\lambda^2+\lambda_R^2k^2}|\boldsymbol{d}_{\perp}|^2\chi_0 \notag \\
	&+\frac{\lambda_R^2k^2}{\lambda^2+\lambda_R^2k^2}(|\boldsymbol{d}_{\parallel}|^2-|\boldsymbol{d}\cdot\hat{\boldsymbol{k}}|^2)(\chi_0+2N\mathcal{C}_0(T,\varepsilon\lambda_Rk_F)) \notag \\
	&\to\begin{cases}
		&\chi_{3u3,3u3}=\frac{\lambda^2}{\lambda^2+\lambda_R^2k_F^2}\chi_0,\quad\boldsymbol{d}=(0,0,1). \\
		&\chi_{3u3e,3u3e}=\frac{\lambda_R^2k_F^2}{\lambda^2+\lambda_R^2k_F^2}(\chi_0+2N\mathcal{C}_0(T,\varepsilon\lambda_Rk_F)), \\
		&\boldsymbol{d}=\sqrt{2}(0,1,0)/\sqrt{2}(1,0,0).
	\end{cases} \label{3u3}
\end{align}
Based on symmetry analysis, the possible combinations between Singlets and Triplet states are: \textcircled{1}$A_1$: 1g1, 1g2, 3u1t, 3u2t, 3gr, 3u3 \textcircled{2}$A_2$: 1u, 3u2r, 3gt \textcircled{3}$E_1$: 1g3, 3u1p, 3u3e \textcircled{4}$E_2$: 3u1e, 3u2e, 3ge, we list out the nonzero superconductivity susceptibilities:
\begin{align}
	\chi_{3gt,1u}&=\chi_{1u,3gt} \notag \\
	&=-\frac{\lambda_Rk_F}{\sqrt{\lambda^2+\lambda_R^2k_F^2}}(\chi_0+2N\mathcal{C}_0(T,\varepsilon\lambda_Rk_F)) \label{3g1u}\\
	\chi_{3u1t,1g1}&=\chi_{1g1,3u1t}=\chi_{3u2t,1g2}=\chi_{1g2,3u2t}\notag \\
	&=-\frac{\lambda_Rk_F}{\sqrt{\lambda^2+\lambda_R^2k_F^2}}\chi_0 \\
	\chi_{3u2r,1u}&=\chi_{1u,3u2r} \notag \\
	&=\frac{\lambda\lambda_Rk_F}{\lambda^2+\lambda_R^2k_F^2}(\chi_0+2N\mathcal{C}_0(T,\varepsilon\lambda_Rk_F)) \\
	\chi_{3u1t,3u3}&=\chi_{3u3,3u1t}=-\chi_{1g2,3gr}=-\chi_{3gr,1g2} \notag \\
	&=\frac{\lambda\lambda_Rk_F}{\lambda^2+\lambda_R^2k_F^2}\chi_0 \\
	\chi_{3u2t,3gr}&=\chi_{3gr,3u2t}=-\chi_{3u3,1g1}=-\chi_{1g1,3u3} \notag \\
	&=\frac{\lambda}{\sqrt{\lambda^2+\lambda_R^2k_F^2}}\chi_0 \\
	\chi_{3u1p,1g3}&=\chi_{1g3,3u1p}=\chi_{3u2r,3gt}=\chi_{3gt,3u2r} \notag \\
	&=-\frac{\lambda}{\sqrt{\lambda^2+\lambda_R^2k_F^2}}(\chi_0+2N\mathcal{C}_0(T,\varepsilon\lambda_Rk_F)) \\
	\chi_{1g3,3u3e}&=\chi_{3u3e,1g3} \notag \\
	&=\mp\frac{\lambda_Rk_F}{\sqrt{\lambda^2+\lambda_R^2k_F^2}}(\chi_0+2N\mathcal{C}_0(T,\varepsilon\lambda_Rk_F)) \\
	\chi_{3u1p,3u3e}&=\chi_{3u3e,3u1p} \notag \\
	&=\pm\frac{\lambda\lambda_Rk_F}{\lambda^2+\lambda_R^2k_F^2}(\chi_0+2N\mathcal{C}_0(T,\varepsilon\lambda_Rk_F)) \\
	\chi_{3u2e,3ge}&=\chi_{3ge,3u2e} \notag \\
	&=\pm\frac{\lambda}{\sqrt{\lambda^2+\lambda_R^2k_F^2}}(\chi_0+N\mathcal{C}_0(T,\varepsilon\lambda_Rk_F)) \label{3u11g3}
\end{align} 
There are five mixing states, a. 1g1, 3u1t, 3u3 ($A_1$); b. 1g2, 3u2t, 3gr ($A_1$); c. 1u, 3u2r, 3gt ($A_2$); d. 1g3, 3u1p, 3u3e ($E_1$); e. 3u2e, 3ge ($E_2$). 

\section{\label{app_phase}Functions for $T_c$}
The critical temperature for each superconducting state is obtained by linear gap function(Eq.(\ref{gap eq}) in the main text). We get functions for $T_c$s to draw phase diagram in pairing interaction parameter space.
\subsection{First Quadrant}
\begin{align}
	&\chi_{0}(T_{a1})=\frac{1}{ v_{0}}\frac{1}{xcos^{2}\phi_{F}+ysin^{2}\phi_{F}+1}, \notag \\
	&(\Delta_{1g1},\Delta_{3u1t},\Delta_{3u3})=\Delta_{a1}(1,-ysin\phi_{F},-xcos\phi_{F}). \label{Ta}\\
	&\chi_{0}(T_{b1})=\frac{1}{ v_{0}}\frac{1}{xsin^{2}\phi_{F}+ycos^{2}\phi_{F}+xy}, \notag \\
	&(\Delta_{1g2},\Delta_{3u2t},\Delta_{3gr})=\Delta_{b1}(xsin\phi_{F},-xy,-ycos\phi_{F}). \\
	&\chi_{0}(T_{c1})+2N\mathcal{C}_{0}(T_{c1},\varepsilon\lambda_{R}k_{F})=\frac{1}{ v_{0}}\frac{1}{y+xycos^{2}\phi_{F}+sin^{2}\phi_{F}}, \notag \\
	&(\Delta_{1u},\Delta_{3u2r},\Delta_{3gt})=\Delta_{c1}(sin\phi_{F},xycos\phi_{F},-y). \label{Tc}\\
	&\chi_{0}(T_{d1})+2N\mathcal{C}_{0}(T_{d1},\varepsilon\lambda_{R}k_{F})=\frac{1}{ v_{0}}\frac{1}{xsin^{2}\phi_{F}+ycos^{2}\phi_{F}+xy}, \notag \\
	&(\Delta_{1g3},\Delta_{3u1p},\Delta_{3u3e})=\Delta_{d1}(xy,-ycos\phi_{F},\mp xsin\phi_{F}). \label{Td} \\
	&\chi_{0}(T_{e1})+N\mathcal{C}_{0}(T_{e1},\varepsilon\lambda_{R}k_{F})\approx\frac{1}{ v_{0}}\frac{1}{(1+x)y}, \notag \\
	&(\Delta_{3u2e},\Delta_{3ge})=\Delta_{e1}(x,\pm1). \label{Te} 
\end{align}
where $\phi_{F}=arctan(\lambda_{R}k_{F}/\lambda)$.
$x,y$ are defined by $x=u_{0}/v_{0}=u/v$ and $y=v/v_{0}=u/u_{0}$.
Here, we assume $\lambda_{R}k_{F}\ll\lambda$, i.e., $\phi_{F}\to0$.
Then, the possible pairings are reduced to only four states: 1g1,
3u2t, 3gr and 3u3. We compare these critical temperatures and
get the phase diagram in the first quadrant of Fig.\ref{fig1}c in the main text. 
\subsection{Second Quadrant}
\begin{align}
	&\chi_0(T_{a2})=\frac{1}{ v_0}\frac{1}{1+ysin^2\phi_F}, \notag \\
	&(\Delta_{1g1},\Delta_{3u1t})=\Delta_{a2}(1,-ysin\phi_F). \\
	&\chi_0(T_{b2})=\frac{1}{ v_0}\frac{1}{ycos^2\phi_F},\quad\Delta_{3gr}=\Delta_{b2}. \\
	&\chi_0(T_{c2})+2N\mathcal{C}_0(T_{c2},\varepsilon\lambda_Rk_F)=\frac{1}{ v_0}\frac{1}{y+sin^2\phi_F}, \notag \\
	&(\Delta_{1u},\Delta_{3gt})=\Delta_{c2}(sin\phi_F,-y). \\
	&\chi_0(T_{d2})+2N\mathcal{C}_0(T_{d2},\varepsilon\lambda_Rk_F)=\frac{1}{ v_0}\frac{1}{ycos^2\phi_F}, \notag \\
	&\Delta_{3u1p}=\Delta_{d2}. \\
	&\chi_0(T_{e2})+N\mathcal{C}_0(T_{e2},\varepsilon\lambda_Rk_F)\approx\frac{1}{ v_0}\frac{1}{y}, \notag \\
	&\Delta_{3ge}=\Delta_{e2}.
\end{align}
\subsection{Third Quadrant}
\begin{align}
	&\chi_0(T_{a3})=\frac{1}{ v_0},\quad\Delta_{1g1}=\Delta_{a3}. \\
	&\chi_0(T_{b3})=\frac{1}{ v_0}\frac{1}{xy},\quad\Delta_{3u2t}=\Delta_{b3}. \\
	&\chi_0(T_{c3})+2N\mathcal{C}_0(T_{c3},\varepsilon\lambda_Rk_F)=\frac{1}{ v_0}\frac{1}{xycos^2\phi_F+sin^2\phi_F}, \notag \\
	&(\Delta_{1u},\Delta_{3u2r})=\Delta_{c3}(sin\phi_F,xycos\phi_F). \\
	&\chi_0(T_{d3})+2N\mathcal{C}_0(T_{d3},\varepsilon\lambda_Rk_F)=\frac{1}{v_0}\frac{1}{xy},\quad\Delta_{1g3}=\Delta_{d3}. \\
	&\chi_0(T_{e3})+N\mathcal{C}_0(T_{e3},\varepsilon\lambda_Rk_F)\approx\frac{1}{v_0}\frac{1}{xy},\quad\Delta_{3u2e}=\Delta_{e3}.
\end{align}
\subsection{Fourth Quadrant}
\begin{align}
	&\chi_0(T_{a4})=\frac{1}{ v_0}\frac{1}{1+xcos^2\phi_F}, \notag \\
	&(\Delta_{1g1},\Delta_{3u3})=\Delta_{a4}(1,-xcos\phi_F). \\
	&\chi_0(T_{b4})=\frac{1}{ v_0}\frac{1}{xsin^2\phi_F},\quad\Delta_{1g2}=\Delta_{b4}. \\
	&\chi_0(T_{c4})+2N\mathcal{C}_0(T_{c4},\varepsilon\lambda_Rk_F)=\frac{1}{ v_0}\frac{1}{sin^2\phi_F},\quad\Delta_{1u}=\Delta_{c4}.\\
	&\chi_0(T_{d4})+2N\mathcal{C}_0(T_{d4},\varepsilon\lambda_Rk_F)=\frac{1}{v_0}\frac{1}{xsin^2\phi_F}, \notag \\
	&\Delta_{3u3e}=\Delta_{d4}.
\end{align}
State e is absent. 

\section{\label{app_field}Superconductivity with magnetic field}
Consider unconventional Rashba model with Zeeman term 
\begin{align}
	H&=H_0+V \notag \\
	H_0&=tk^2\sigma^0\tau^0+\lambda\sigma^3\tau^2 \notag \\
	V&=-\lambda_R(k_y\sigma^1-k_x\sigma^2)(\tau^0+\varepsilon\tau^1)+h\sigma^3\tau^0 \label{H0h}
\end{align}
The symmetry remained in this Hamiltonian is group $SO(2)$, since Zeeman term breaks mirror symmetry. All IRs for $SO(2)$ are one dimensional with characters in form of $e^{im\theta}(m=0,\pm1,\pm2...)$. Thus the pairings considered before are reduced to IR $A$($m=0$), IR $B_\pm$($|m|=1$) and IR $C_\pm$($|m|=2$). We list the pairings in a new table(Table \ref{pairh}) according to group $SO(2)$. 
\begin{table}[ptb]
	\caption{$\phi_{g/u}$ and $\boldsymbol{d_{g/u}}$ are scalar
		and vector functions of $k$, respectively with g and u labeling even
		and odd under transformation $\bm{k}\to-\bm{k}$. The second column
		lists $\phi_{g/u}$ and $\boldsymbol{d_{g/u}}$ for simplist pairings to the lowest order of $k$, where $\hat{k}_{x}^{2}+\hat{k}_{y}^{2}=1$
		is normalized. The corresponding IRs for $SO(2)$ are listed in
		the third column, the same format of fonts denotes these pairings
		can be mixed. In the last column, we follow labels in the main text. }
	\label{pairh}
	\begin{ruledtabular}
		\begin{tabular}{cccc}
			Pairing form  & $\phi_{g/u}$ / $\boldsymbol{d_{g/u}}$   & IRs for $SO(2)$  & Label\tabularnewline
			\hline 
			$i\phi_{g}\sigma^{2}\tau^{0}$  & $1$  & $A$  & 1g1 \tabularnewline
			$i\phi_{u}\sigma^{2}\tau^{2}$  & $\hat{k}_{x}\pm i\hat{k}_{y}$  & $\mathcal{B_{\pm}}$  & 1g3 \tabularnewline
			$i(\boldsymbol{d_{u}}\cdot\boldsymbol{\sigma})\sigma^{2}\tau^{0}$  & $(0,0,\hat{k}_{x}\pm i\hat{k}_{y})$  & $\mathcal{B}_{\pm}$  & 3u1p \tabularnewline
			$i(\boldsymbol{d_{u}}\cdot\boldsymbol{\sigma})\sigma^{2}\tau^{1}$  & $(\hat{k}_{x},\hat{k}_{y},0)$  & $A$  & 3u2r \tabularnewline
			& $(-\hat{k}_{y},\hat{k}_{x},0)$  & $\mathbf{A}$  & 3u2t \tabularnewline
			& $(\hat{k}_x\pm i\hat{k}_y)(1,\pm i,0)/\sqrt{2}$ & $\mathcal{C}_{\pm}$ & 3u2e\tabularnewline
			$i(\boldsymbol{d_{u}}\cdot\boldsymbol{\sigma})\sigma^{2}\tau^{3}$  & $(\hat{k}_{x},\hat{k}_{y},0)$  & $\mathbf{A}$  & 3gr \tabularnewline
			& $(-\hat{k}_{y},\hat{k}_{x},0)$  & $A$  & 3gt \tabularnewline
			& $(\hat{k}_x\pm i\hat{k}_y)(1,\pm i,0)/\sqrt{2}$ & $\mathcal{C}_{\pm}$ & 3ge\tabularnewline
			$i(\boldsymbol{d_{g}}\cdot\boldsymbol{\sigma})\sigma^{2}\tau^{2}$  & $(0,0,1)$  & $A$  & 3u3 \tabularnewline
		\end{tabular}
	\end{ruledtabular}
\end{table}
Solution of the revised Hamiltonian is carried out by perturbation method
\begin{align}
	E_{\alpha\beta}=tk^2+\alpha\lambda+\beta\sqrt{\varepsilon^2\lambda_R^2 k^2+h^2}, \quad \alpha,\beta\in\{+,-\}
\end{align}
The unitary matrix to approximately diagonalize $H$ is
\begin{equation}
	U=\frac{1}{\sqrt{2}}\begin{pmatrix}
		-icos\frac{\varphi}{2} & -isin\frac{\varphi}{2} & icos\frac{\varphi}{2} & isin\frac{\varphi}{2} \\
		ie^{i\theta}sin\frac{\varphi}{2} & -ie^{i\theta}cos\frac{\varphi}{2} & ie^{i\theta}sin\frac{\varphi}{2} & -ie^{i\theta}cos\frac{\varphi}{2} \\
		cos\frac{\varphi}{2} & sin\frac{\varphi}{2} & cos\frac{\varphi}{2} & sin\frac{\varphi}{2} \\
		e^{i\theta}sin\frac{\varphi}{2} & -e^{i\theta}cos\frac{\varphi}{2} & -e^{i\theta}sin\frac{\varphi}{2} & e^{i\theta}cos\frac{\varphi}{2}
	\end{pmatrix}
\end{equation}
where $\varphi=arctan(\varepsilon\lambda_R k/h)$. Thus Green function for electron is
\begin{align}
	&G^0(\boldsymbol{k},i\omega_n)=G_{++++}\sigma^0\tau^0+G_{++--}\sigma^3\tau^2 \notag \\
	&+G_{+-+-}\left(\frac{\varepsilon\lambda_Rk}{\sqrt{\varepsilon^2\lambda_R^2k^2+h^2}}|\boldsymbol{k}\times\boldsymbol{\sigma}|_z\tau^1+\frac{h}{\sqrt{\varepsilon^2\lambda_R^2k^2+h^2}}\sigma^3\tau^0\right) \notag \\
	&+G_{+--+}\left(\frac{h}{\sqrt{\varepsilon^2\lambda_R^2k^2+h^2}}\sigma^0\tau^2-\frac{\varepsilon\lambda_Rk}{\sqrt{\varepsilon^2\lambda_R^2k^2+h^2}}(\boldsymbol{k}\cdot\boldsymbol{\sigma})\tau^3\right)
\end{align}
where $G_{\alpha\beta\gamma\delta}=(\alpha \tilde{G}_{++}+\beta \tilde{G}_{+-}+\gamma \tilde{G}_{-+}+\delta \tilde{G}_{--})/4$, $ \tilde{G}_{\alpha\beta}(\boldsymbol{k},i\omega_n)=1/(i\omega_n-E_{\alpha\beta})$. And we find  $g_{\mu\nu\rho\sigma}^{\alpha\beta\gamma\delta}$s are different from the ones in Appendix \ref{app_susc}. 
\begin{align}
	g_{++++}^{++++}&=g_{++--}^{++--}=-g_{++--}^{++++}=-g_{++++}^{++--} \notag \\
	&=\frac{1}{8}\chi_0+\frac{1}{8}N\mathcal{C}_0(T,\sqrt{\varepsilon^2\lambda_R^2k_F^2+h^2}) \notag \\
	g_{+-+-}^{+-+-}&=g_{+--+}^{+--+}=-g_{+--+}^{+-+-}=-g_{+-+-}^{+--+} \notag \\
	&=-\frac{1}{8}N\mathcal{C}_0(T,\sqrt{\varepsilon^2\lambda_R^2k_F^2+h^2}) \notag
\end{align}
The self-superconductivity-susceptibilities are
\begin{align}
	\chi_{1g1,1g1}&=\chi_{3u3,3u3} \notag \\
	&=\chi_0+\frac{h^2}{\varepsilon^2\lambda_R^2k_F^2+h^2}2N\mathcal{C}_0(T,\sqrt{\varepsilon^2\lambda_R^2k_F^2+h^2})  \\
	\chi_{3u2r,3u2r}&=\chi_{3gt,3gt} \notag \\
	&=\chi_0+\frac{\varepsilon^2\lambda_R^2k_F^2}{\varepsilon^2\lambda_R^2k_F^2+h^2}2N\mathcal{C}_0(T,\sqrt{\varepsilon^2\lambda_R^2k_F^2+h^2}) \\
	\chi_{3u2t,3u2t}&=\chi_{3gr,3gr}=\chi_0 \\
	\chi_{1g3,1g3}&=\chi_{3u1p,3u1p}=\chi_0+2N\mathcal{C}_0(T,\sqrt{\varepsilon^2\lambda_R^2k_F^2+h^2}) \\
	\chi_{3u2e,3u2e}&=\chi_{3ge,3ge} \notag \\
	&=\chi_0+\frac{\varepsilon^2\lambda_R^2k_F^2}{\varepsilon^2\lambda_R^2k_F^2+h^2}N\mathcal{C}_0(T,\sqrt{\varepsilon^2\lambda_R^2k_F^2+h^2})
\end{align}
The nonzero inter-superconductivity-susceptibilities of possible mixing states are
\begin{align}
	\chi_{1g1,3u2r}&=\chi_{3u3,3gt}=-\chi_{3u2r,1g1}=-\chi_{3gt,3u3} \notag \\
	&=-i\frac{h\varepsilon\lambda_Rk_F}{\varepsilon^2\lambda_R^2k_F^2+h^2}2N\mathcal{C}_0(T,\sqrt{\varepsilon^2\lambda_R^2k_F^2+h^2}) \\
	\chi_{1g1,3gt}&=\chi_{3u3,3u2r}=-\chi_{3gt,1g1}=-\chi_{3u2r,3u3} \notag \\
	&=i\frac{h\varepsilon\lambda_Rk_F}{\varepsilon^2\lambda_R^2k_F^2+h^2}2N\mathcal{C}_0(T,\sqrt{\varepsilon^2\lambda_R^2k_F^2+h^2}) \\
	\chi_{1g1,3u3}&=\chi_{3u3,1g1} \notag \\
	&=-[\chi_0+\frac{h^2}{\varepsilon^2\lambda_R^2k_F^2+h^2}2N\mathcal{C}_0(T,\sqrt{\varepsilon^2\lambda_R^2k_F^2+h^2})] \\
	\chi_{3u2r,3gt}&=\chi_{3gt,3u2r} \notag \\
	&=-[\chi_0+\frac{\varepsilon^2\lambda_R^2k_F^2}{\varepsilon^2\lambda_R^2k_F^2+h^2}2N\mathcal{C}_0(T,\sqrt{\varepsilon^2\lambda_R^2k_F^2+h^2})] \\
	\chi_{3u2t,3gr}&=\chi_{3gr,3u2t}=\chi_0 \\
	\chi_{1g3,3u1p}&=\chi_{3u1p,1g3}=-[\chi_0+2N\mathcal{C}_0(T,\sqrt{\varepsilon^2\lambda_R^2k_F^2+h^2})] \\
	\chi_{3u2e,3ge}&=-\chi_{3ge,3u2e} \notag \\
	&=-i[\chi_0+\frac{\varepsilon^2\lambda_R^2k_F^2}{\varepsilon^2\lambda_R^2k_F^2+h^2}N\mathcal{C}_0(T,\sqrt{\varepsilon^2\lambda_R^2k_F^2+h^2})]
\end{align}
There remain four mixing states, $\mathscr{A}$. 1g1, 3u2r, 3gt, 3u3; $\mathscr{B}$. 3u2t, 3gr; $\mathscr{D}$. 1g3, 3u1p; $\mathscr{E}$. 3u2e, 3ge. States $\mathscr{A}$, $\mathscr{B}$ are $A$ representations and states $\mathscr{D}$, $\mathscr{E}$ are $B$ and $C$ representations for group $SO(2)$. We notice state $\mathscr{A}$ is the combination of states a and c for zero field case. 
\subsection{First Quadrant}
In the first quadrant, we get the critical temperatures
\begin{align}
	&\chi_0(T_{\mathscr{A}1})+\mathcal{C}_0=\frac{1}{2v_0(1+x)y}\{1+y \notag \\
	&-\sqrt{4v_0^2(1+x)^2y^2\mathcal{C}_0^2+4v_0(1+x)(1-y)cos2\varphi\mathcal{C}_0+(1-y)^2}\}. \label{ha}\\
	&\chi_0(T_{\mathscr{B}1})=\frac{1}{v_0(1+x)y}. \\
	&\chi_0(T_{\mathscr{D}1})+2\mathcal{C}_0=\frac{1}{v_0(1+x)y}. \label{hc} \\
	&\chi_0(T_{\mathscr{E}1})+\frac{\varepsilon^2\lambda_R^2k_F^2}{\varepsilon^2\lambda_R^2k_F^2+h^2}N\mathcal{C}_0=\frac{1}{v_0(1+x)y}.
\end{align}
where $\mathcal{C}_0=N\mathcal{C}_0(T,\lambda_Rk_F,h)$. We find the state $\mathscr{B}$ is not affected by the magnetic field, while states $\mathscr{A}$, $\mathscr{D}$ and $\mathscr{E}$ are affected differently. To investigate the difference of $\mathscr{A}$ and $\mathscr{D}$, one could make a subtraction between Eq.(\ref{ha},\ref{hc})
\begin{align}
	&2ln(\frac{T_{\mathscr{D}1}}{T_{\mathscr{A}1}})=\mathcal{C}_0-\frac{1}{2v_0(1+x)y}\{1-y \notag \\
	&+\sqrt{4v_0^2(1+x)^2y^2\mathcal{C}_0^2+4v_0(1+x)(1-y)cos2\varphi\mathcal{C}_0+(1-y)^2}\}
\end{align}
If we want $T_{\mathscr{D}1}>T_{\mathscr{A}1}$, parameters should obey $\{1/[1+2v_0(1+x)\mathcal{C}_0],-cos2\varphi\}_{max}<y<1$, however the situation could never be satisfied since $\mathcal{C}_0<0$. Therefore state $\mathscr{D}$ is not affected less by magnetic field than state $\mathscr{A}$. The similar results could be obtained for the other quadrants.
\subsection{Second Quadrant}
\begin{align}
	&\chi_0(T_{\mathscr{A}2})+\mathcal{C}_0=\frac{1}{2v_0y}\{1+y \notag \\
	&-\sqrt{4v_0^2y^2\mathcal{C}_0^2+4v_0y(1-y)cos2\varphi\mathcal{C}_0+(1-y)^2}\}, \notag \\
	&(\Delta_{1g1},\Delta_{3gt}). \label{h2}\\
	&\chi_0(T_{\mathscr{B}2})=\frac{1}{v_0y},\quad\Delta_{3gr}. \\
	&\chi_0(T_{\mathscr{D}2})+2\mathcal{C}_0=\frac{1}{v_0y},\quad\Delta_{3u1p}. \\
	&\chi_0(T_{\mathscr{E}2})+\frac{\varepsilon^2\lambda_R^2k_F^2}{\varepsilon^2\lambda_R^2k_F^2+h^2}N\mathcal{C}_0=\frac{1}{v_0y}, \quad\Delta_{3ge}.
\end{align}
\subsection{Third Quadrant}
\begin{align}
	&\chi_0(T_{\mathscr{A}3})+\mathcal{C}_0=\frac{1}{2v_0xy}\{1+xy \notag \\
	&-\sqrt{4v_0^2x^2y^2\mathcal{C}_0^2+4v_0xy(1-xy)cos2\varphi\mathcal{C}_0+(1-xy)^2}\}, \notag \\
	&(\Delta_{1g1},\Delta_{3u2r}). \\
	&\chi_0(T_{\mathscr{B}3})=\frac{1}{v_0xy},\quad\Delta_{3u2t}. \\
	&\chi_0(T_{\mathscr{D}3})+2\mathcal{C}_0=\frac{1}{v_0xy},\quad\Delta_{1g3}. \\
	&\chi_0(T_{\mathscr{E}3})+\frac{\varepsilon^2\lambda_R^2k_F^2}{\varepsilon^2\lambda_R^2k_F^2+h^2}N\mathcal{C}_0=\frac{1}{v_0xy}, \quad\Delta_{3u2e}.
\end{align}
\subsection{Fourth Quadrant}
\begin{align}
	\chi_0(T_{\mathscr{A}4})&+2cos^2\varphi\mathcal{C}_0=\frac{1}{v_0(1+x)},\quad(\Delta_{1g1},\Delta_{3u3}). \label{h4}
\end{align}
States $\mathscr{B}$, $\mathscr{D}$ and $\mathscr{E}$ are absent in this situation.
\section{\label{app_tunnel}Tunneling Process}
To get the information of the interaction parameters, we find a way to detect which one of the two p-wave triplet state is leading. The eigenstates of quasi-particles are solutions of the $8\times8$ Bogoliubov-de Gennes(BdG) Hamiltonian(\ref{bdg}) in Numbu space $(c_{\boldsymbol{k},1\uparrow}^{\dagger},c_{\boldsymbol{k},1\downarrow}^{\dagger},c_{\boldsymbol{k},2\uparrow}^{\dagger},c_{\boldsymbol{k},2\downarrow}^{\dagger},c_{-\boldsymbol{k},1\uparrow},c_{-\boldsymbol{k},1\downarrow},c_{-\boldsymbol{k},2\uparrow},c_{-\boldsymbol{k},2\downarrow})$, which have tedious spinor that make the tunneling problem difficult to solve, especially for the states 3u3 and 3gr. 
\begin{equation}
	H_{BdG}(\boldsymbol{k})=\begin{pmatrix}
		H_0(\boldsymbol{k})&\Delta(\boldsymbol{k})\\\Delta^{\dagger}(\boldsymbol{k})&-H_0^{T}(-\boldsymbol{k}) \label{bdg}
	\end{pmatrix}
\end{equation}
$\Delta(\boldsymbol{k})$ is constructed by order parameter $\Delta$ and specific pairing form in Table.\uppercase\expandafter{\romannumeral1} of the main text. We assume order parameters are real for convenience, since the calculation results must be independent of phases. Fortunately, the upper two bands affect little to the lower bands as $\lambda_Rk_F\ll\lambda$, therefore we would make a downfolding approach and get an effective $4\times4$ BdG Hamitonian. We firstly make a rotation in the spinor space. 
\begin{align}
	\tilde{H}_{BdG}&=\begin{pmatrix}
		U_r^{\dagger}&\\&U_r^{T}
	\end{pmatrix}H_{BdG}\begin{pmatrix}
		U_r&\\&U_r^{*}
	\end{pmatrix} \notag \\
    &=\begin{pmatrix}
		\tilde{H}_0(\boldsymbol{k})&\tilde{\Delta}(\boldsymbol{k})\\\tilde{\Delta}^{\dagger}(\boldsymbol{k})&-\tilde{H}_0^{T}(-\boldsymbol{k})
	\end{pmatrix}
\end{align}
Here $U_r=(\sigma^0\tau^0+i\sigma^3\tau^1)/\sqrt{2}$, and thus parameter $\lambda$ goes into diagonal elements of the matrix. The block elements are listed below, superconducting states for 1g1, 3u3, 3u2t and 3gr are considered. 
\begin{align}
	\tilde{H}_0(\boldsymbol{k})&=\xi_{\boldsymbol{k}}\sigma^0\tau^0-\lambda_{R}(k_x\sigma^1+k_y\sigma^2)(\varepsilon\tau^0+\tau^1)+\lambda\sigma^0\tau^3 \\
	\tilde{\Delta}^{1g1}(\boldsymbol{k})&=i\Delta\sigma^2\tau^0 \\
	\tilde{\Delta}^{3u3}(\boldsymbol{k})&=-i\Delta\sigma^2\tau^3 \\
	\tilde{\Delta}^{3u2t}(\boldsymbol{k})&=\Delta(\hat{k}_x\sigma^3\tau^0-i\hat{k}_y\sigma^0\tau^0) \\
	\tilde{\Delta}^{3gr}(\boldsymbol{k})&=\Delta(-\hat{k}_x\sigma^3\tau^3+i\hat{k}_y\sigma^0\tau^3)
\end{align}
In order to get results in a compact form, we choose $\Delta\to-\Delta$ for state 3u3, comparing with the other three states. After rearranging the basis in $(c_{\boldsymbol{k},+\uparrow}^{\dagger},c_{\boldsymbol{k},+\downarrow}^{\dagger},c_{-\boldsymbol{k},+\uparrow},c_{-\boldsymbol{k},+\downarrow},c_{\boldsymbol{k},-\uparrow}^{\dagger},c_{\boldsymbol{k},-\downarrow}^{\dagger},c_{-\boldsymbol{k},-\uparrow},c_{-\boldsymbol{k},-\downarrow})$, where $+$ and $-$ denote combinations of orbitals that have diagonal elements $\xi_{\boldsymbol{k}}+\lambda$ and $\xi_{\boldsymbol{k}}-\lambda$, we get the Hamiltonian in form of
\begin{equation}
	\tilde{H}_{BdG}^{r}=\begin{pmatrix}
		\hat{h}^{+}+\hat{\Delta}^{+}&T\\T^{\dagger}&\hat{h}^{-}+\hat{\Delta}^{-}
	\end{pmatrix} \label{rbdg}
\end{equation}
The elements in Eq.(\ref{rbdg}) are
\begin{align}
	\hat{h}^{\pm}&=\hat{h}_{0}^{\pm}+\hat{h}_{R},\quad\hat{h}_{0}^{\pm}=(\xi_{\boldsymbol{k}}\pm\lambda)\sigma^0\eta^3, \notag \\
	\hat{h}_{R}&=-\varepsilon\lambda_{R}(k_x\sigma^1\eta^0+k_y\sigma^2\eta^3). \\
	\hat{\Delta}^{+}_{1g1}&=\hat{\Delta}^{-}_{1g1}=-\hat{\Delta}^{+}_{3u3}=\hat{\Delta}^{-}_{3u3}=-\Delta\sigma^2\eta^2 \\
	\hat{\Delta}^{+}_{3u2t}&=\hat{\Delta}^{-}_{3u2t}=-\hat{\Delta}^{+}_{3gr}=\hat{\Delta}^{-}_{3gr} \notag \\
	&=\Delta(\hat{k}_x\sigma^3\eta^1+\hat{k}_y\sigma^0\eta^2) \\
	T&=-\lambda_{R}(k_x\sigma^1\eta^0+k_y\sigma^2\eta^3)
\end{align}
Pauli matrices $(\eta^0,\boldsymbol{\eta})$ span the particle-hole space. Then use the downfolding approach, and find the effective BdG Hamiltonian for the lower bands. 
\begin{align}
	\hat{H}_{BdG}^{eff}&\approx\hat{h}^{-}+\hat{\Delta}^{-}-T(\hat{h}^{+}+\hat{\Delta}^{+})^{-1}T^{\dagger} \notag \\
	&\approx\hat{h}^{-}+\hat{\Delta}^{-}+\delta \hat{h} \label{eff}
\end{align}
$\delta\hat{h}$ describes the influence from the upper bands, which is different by states. For s-wave states and p-wave states respectively, 
\begin{align}
	\delta
	\hat{h}_{s}^{\nu}&=-\frac{\lambda_{R}^2k^2}{2\lambda}\left[(u_{\lambda}^2-v_{\lambda}^2)\sigma^0\eta^3+2\nu u_{\lambda}v_{\lambda}\sigma^2\eta^2\right] \\ \delta\hat{h}_{p}^{\nu}&=\frac{\lambda_{R}^2k}{2\lambda}\left[(v_{\lambda}^2-u_{\lambda}^2)k\sigma^0\eta^3 +2\nu u_{\lambda}v_{\lambda}(k_x\sigma^3\eta^1+k_y\sigma^0\eta^2)\right]
\end{align}
$\nu=+1(-1)$ in $\hat{h}_{s}^{\nu}$ denotes states 1g1(3u3), while $\nu=+1(-1)$ in $\hat{h}_{p}^{\nu}$ denotes states 3u2t(3gr). $u(v)_{\lambda}^2=(\sqrt{(2\lambda)^2+\Delta^2}+(-)2\lambda)/2\sqrt{(2\lambda)^2+\Delta^2}$. Positive eigenenergys and corresponding eigenstates of the effective Hamiltonian(\ref{eff}) have compact form. 
\begin{align}
	&E_{s,\pm}^{\nu}=E_{p,\pm}^{\nu}=E_{\pm}^{\nu} \notag \\
	&=\sqrt{(\xi_{\boldsymbol{k}}-\lambda\pm\varepsilon\lambda_{R}k-\epsilon_0cos2\gamma)^2+(\Delta+\nu\epsilon_0sin2\gamma)^2} \\
	&\psi_{s,\pm}^{\nu}=(u_{\nu},\mp e^{i\theta}u_{\nu},\pm e^{i\theta}v_{\nu},v_{\nu}), \notag \\
	&\psi_{p,\pm}^{\nu}=(\pm u_{\nu},-e^{i\theta}u_{\nu},\pm e^{i\theta}v_{\nu},v_{\nu})
\end{align}
Here $\gamma=arctan(v_{\lambda}/u_{\lambda})$, $\epsilon_0=\lambda_{R}^2k^2/2\lambda$, in tunneling problem $u(v)_{\nu}^2=(E+(-)\sqrt{E^2-(\Delta+\nu\epsilon_0sin2\gamma)^2})/2E$ for the given energy $E$. 

Then we solve Eq.(\ref{junction}) in the main text. Supposing the condition $(\lambda_Rk_F,\sqrt{E^2-\Delta^2})\ll\mu$ as $\mu$ is the same order comparing with $\lambda$, wave vectors of the normal metal electron and BdG quasi-particle states are approximately the same and equal to $k_F=\sqrt{2m\mu}/\hbar$. Since the unconventional Rashba bands have infinitesimal spin structures as $\lambda_Rk_F\ll\lambda$, conductance would not be differed by spin, thus we arbitrarily choose a spin-up incident electron. The wave function for region $x<0$ is 
\begin{align}
	&\psi_L(x)=e^{ik_yy}[e^{ik_xx}\begin{pmatrix}
		1\\
		0\\
		0\\
		0
	\end{pmatrix}+b_{11}e^{-ik_xx}\begin{pmatrix}
		1\\
		0\\
		0\\
		0
	\end{pmatrix}
	 \notag \\
	&+b_{12}e^{-ik_xx}\begin{pmatrix}
		0\\
		1\\
		0\\
		0
	\end{pmatrix}+a_{11}e^{ik_xx}\begin{pmatrix}
		0\\
		0\\
		1\\
		0
	\end{pmatrix}+a_{12}e^{ik_xx}\begin{pmatrix}
		0\\
		0\\
		0\\
		1
	\end{pmatrix}] \label{psil}
\end{align}
For 3u2t or 3gr, the wave function for region $x>0$ is 
\begin{align}
	&\psi_R^{\nu}(x)=\frac{1}{\sqrt{2}}e^{ik_yy}[c_{11}e^{ik_xx}\begin{pmatrix}
		u_{\nu}\\
		-e^{i\theta}u_{\nu} \\
		e^{i\theta}v_{\nu} \\
		v_{\nu}
	\end{pmatrix} \notag \\
    &+c_{12}e^{ik_xx}\begin{pmatrix}
		-u_{\nu}\\
		-e^{i\theta}u_{\nu} \\
		-e^{i\theta}v_{\nu} \\
		v_{\nu}
	\end{pmatrix}+d_{11}e^{-ik_xx}\begin{pmatrix}
		v_{\nu}\\
		e^{-i\theta}v_{\nu} \\
		-e^{-i\theta}u_{\nu} \\
		u_{\nu}
	\end{pmatrix} \notag \\
    &+d_{12}e^{-ik_xx}\begin{pmatrix}
		-v_{\nu}\\
		e^{-i\theta}v_{\nu} \\
		e^{-i\theta}u_{\nu} \\
		u_{\nu}
	\end{pmatrix}]\label{psir}
\end{align}
Terms after coefficients '$c$'s are electron-like quasiparticle transmissions and terms after '$d$'s are hole-like quasiparticle transmissions. The coefficients a,b,c,d are determined by equations:
\begin{align}
	\psi_L(0^-)&=\psi_R(0^+) \\
	\frac{d\psi_R}{dx}|_{0^+}&-\frac{d\psi_L}{dx}|_{0^-}=\frac{2mU_0}{\hbar^2}\psi(0)
\end{align}
The results are
\begin{align}
	a_{11}&=\frac{4e^{i\theta}u_{\nu}v_{\nu}}{(4+Z^2)u_{\nu}^2+e^{2i\theta}Z^2v_{\nu}^2} \\
	a_{12}&=0 \\
	b_{11}&=-\frac{(u_{\nu}^2+e^{2i\theta}v_{\nu}^2)Z(Z+2i)}{(4+Z^2)u_{\nu}^2+e^{2i\theta}Z^2v_{\nu}^2} \\
	b_{12}&=0 \\
	c_{11}&=-\frac{i\sqrt{2}u_{\nu}(Z+2i)}{(4+Z^2)u_{\nu}^2+e^{2i\theta}Z^2v_{\nu}^2} \\
	c_{12}&=\frac{i\sqrt{2}u_{\nu}(Z+2i)}{(4+Z^2)u_{\nu}^2+e^{2i\theta}Z^2v_{\nu}^2} \\
	d_{11}&=-\frac{i\sqrt{2}e^{2i\theta}v_{\nu}Z}{(4+Z^2)u_{\nu}^2+e^{2i\theta}Z^2v_{\nu}^2} \\
	d_{12}&=\frac{i\sqrt{2}e^{2i\theta}v_{\nu}Z}{(4+Z^2)u_{\nu}^2+e^{2i\theta}Z^2v_{\nu}^2}
\end{align}
We also get the coefficients for 1g1 and 3u3
\begin{align}
	a_{11}&=0 \\
	a_{12}&=\frac{4u_{\nu}v_{\nu}}{(4+Z^2)u_{\nu}^2-Z^2v_{\nu}^2} \\
	b_{11}&=-\frac{(u_{\nu}^2-v_{\nu}^2)Z(Z+2i)}{(4+Z^2)u_{\nu}^2-Z^2v_{\nu}^2} \\
	b_{12}&=0 \\
	c_{11}&=c_{12}=-\frac{i\sqrt{2}u_{\nu}(Z+2i)}{(4+Z^2)u_{\nu}^2-Z^2v_{\nu}^2} \\
	d_{11}&=d_{12}=\frac{i\sqrt{2}v_{\nu}Z}{(4+Z^2)u_{\nu}^2-Z^2v_{\nu}^2}
\end{align}
where $Z=2mU_0/\hbar^2k_F$. The tunneling conductance normalized by normal state is
\begin{equation}
	G=\frac{1}{2}\int_{-\pi/2}^{\pi/2}d\theta cos\theta\left(1+|a_{11}|^2+|a_{12}|^2-|b_{11}|^2-|b_{12}|^2\right)
\end{equation}
Then we get the conductances for s-wave states and p-wave states
\begin{align}
	G_{s}^{\nu}&=\frac{4}{(4+Z^2)u_{\nu}^2-Z^2v_{\nu}^2} \\
	G_{p}^{\nu}&=2\int_{-\pi/2}^{\pi/2}d\theta cos\theta \notag \\
	&\times\frac{(Z^2+4)u_{\nu}^2-Z^2v_{\nu}^2}{[(4+Z^2)u_{\nu}^2+Z^2v_{\nu}^2cos(2\theta)]^2+[Z^2v_{\nu}sin(2\theta)]^2}
\end{align}
together with $u(v)_{\nu}^2\approx(E+(-)\sqrt{E^2-\Delta^2(1+\nu(\lambda_Rk_F/2\lambda)^2)^2})/2E$, we plot the conductances as functions of $Z$ at a selected tunneling energy $E$ in Fig. \ref{fig-property}(d) of the main text. Notice there is a term with $\sqrt{E^2-\Delta^2(1+\nu(\lambda_Rk_F/2\lambda)^2)^2}$ in $u(v)_{\nu}$, we should be careful when letting $E\to\Delta$ to avoid in gap tunneling. Further, if we consider the zero barrier limit $Z=0$, we have
\begin{align}
	G_{s}^{\nu}=G_{p}^{\nu}=G^{\nu}=\frac{1}{u_{\nu}^2}
\end{align}
The difference between the two conductances is calculated as
\begin{align}
	\delta G&=\frac{1}{u_{1}^2}-\frac{1}{u_{2}^2} \notag \\
	&\approx 2(v_{1}^2-v_{2}^2) \notag \\
	&\approx2\left(\frac{\lambda_Rk_F}{2\lambda}\right)^2\frac{\Delta^2}{E\sqrt{E^2-\Delta^2}}
\end{align}
where we have used the approximation that $(\lambda_Rk_F/2\lambda)^2\Delta^2/(E^2-\Delta^2)$ is a small value, which is satisfied by the parameters we use in Fig. \ref{fig-property}(d).

\end{document}